\documentclass[aps,prb,twocolumn,showpacs,amsmath,amssymb]{revtex4}
\usepackage{epsfig}
\usepackage{graphicx}
\usepackage{bm}
\usepackage[dvips]{color}
\usepackage{braket}
\usepackage{sidecap}
\newcommand{\phiv}{\mbox{\boldmath$\phi$}}

\newcommand{\ddphiv}{\mbox{\boldmath$\ddot{\phi}$}}
\newcommand{\varphiv}{\mbox{\boldmath$\varphi$}}

\newcommand{\ddvarphiv}{\mbox{\boldmath$\ddot{\varphi}$}}
\newcommand{\bin}[2]{\left(\begin{array}{c}
#1\\
#2
\end{array}\right)}

\date{\today}

\DeclareMathOperator{\tr}{tr}

\begin{document}

\title{Microwave quantum optics and electron transport through a metallic dot
  strongly coupled to a transmission line cavity.}

\author{C. Bergenfeldt and P. Samuelsson} \affiliation{Division of
Mathematical Physics, Lund University, Box 118, S-221 00 Lund, Sweden}

\begin{abstract} 
  We investigate theoretically the properties of the photon state and
  the electronic transport in a system consisting of a metallic
  quantum dot strongly coupled to a superconducting microwave
  transmission line cavity.  Within the framework of circuit quantum
  electrodynamics we derive a Hamiltonian for arbitrary strong capacitive
  coupling between the dot and the cavity. The dynamics of the system
  is described by a quantum master equation, accounting for the
  electronic transport as well as the coherent, non-equilibrium
  properties of the photon state. The photon state is investigated,
  focusing on, for a single active mode, signatures of microwave
  polaron formation and the effects of a non-equilibrium photon
  distribution. For two active photon modes, intra mode conversion and
  polaron coherences are investigated. For the electronic transport,
  electrical current and noise through the dot and the
  influence of the photon state on the transport properties are at the
  focus. We identify clear transport signatures due to the
  non-equilibrium photon population, in particular the emergence of
  superpoissonian shot-noise at ultrastrong dot-cavity couplings.
\end{abstract}

\maketitle

\section{Introduction}
The field of circuit Quantum Electro Dynamics (QED) has over the last
decade emerged as an on-chip version of cavity QED. In circuit QED the
interaction between solid-state quantum systems and high-quality
on-chip circuit elements is investigated.  The pioneering works of the
Yale group proposed \cite{Blais04} and demonstrated \cite{Wall2004}
strong coupling between a superconducting qubit and a microwave
transmission line resonator. This opened up for an impressive
development in the field of quantum information processing with
superconducting circuits, \cite{Schoel2008} with a number of key
experiments demonstrating e.g. long distance qubit state
transfer,\cite{Sillanpaa07,Majer07} controllable multi-qubit
entanglement\cite{DiCarlo09} and the execution of basic quantum
algorithms. \cite{DiCarlo10} Recently also nanoscale qubits, based on
e.g. semiconductor nanowires or carbon nanotubes, coupled to
transmission lines, have received increasing
attention.\cite{Childress04,Burkard06,Trif08,Guo08,Lambert09,Cottet10,Cottet102}
A parallel development concerned the possibilities to perform
fundamental quantum optics experiments with microwave photons in
cavities. Experiments on microwave quantum optics range from arbitrary
photon state preparation \cite{hofheinz09} and entanglement of cavity
photons\cite{Wang11} to single photon generation, \cite{Houck07}
microwave lasing\cite{Astafaiev07} and fast tuning of cavity photon
properties.\cite{Sandberg08,Wilson10}
\\
\\
An important recent development is the efforts to reach the
ultrastrong coupling regime, where the strength of the coupling
between the qubit and the cavity becomes comparable to the frequency
of the fundamental cavity mode. In this regime the Jaynes-Cummings
model breaks down and new physical effects become important. Recent
experiments \cite{Niemczyk10,Mooij10,Mooij102} with flux qubits
directly coupled to a superconductor transmission line cavity
demonstrated couplings of the order of ten percent of the resonator
frequency. These findings spurred a number of theoretical works on
microwave quantum optics in the ultrastrong regime, see
e.g. Refs. \onlinecite{Casanova10,Ashhab10,Hausinger10}.
\\
\\
Lately also systems with mesoscopic or nanoscale conductors, such as
Josephson junctions, \cite{Basset10,Pashkin11,hofheinz11}
superconducting single electrons
transistors\cite{Astafaiev07,Marthaler08,Pashkin11} and quantum dots,
\cite{Frey11,Delbecq11,Frey112,Jin11} inserted into microwave
cavities have been investigated. In particular, the spectral
properties of microwaves emitted from a Josephson junction in the
dynamical Coulomb blockade regime were investigated in
Ref. \onlinecite{hofheinz11}. Also, in a number of very recent experiments,
single \cite{Frey11,Delbecq11} and double \cite{Frey112} quantum dots
were coupled to external leads and the electronic transport was
investigated via the scattering properties of injected
microwaves. Moreover, microwave lasing with population inversion
caused by electron tunneling through a superconducting single
electron transistor was demonstrated experimentally\cite{Astafaiev07}
and investigated theoretically.\cite{rodrigues07,Marthaler08} These experimental
achievements open up for a detailed investigation of the interplay of
transport electrons and individual cavity photons. Of particular
interest is the {\it strong coupling} regime, where the rate for
tunnel induced photon excitation (and de-excitation) is much larger
than the intrinsic cavity photon decay rate. In this regime the photon
distribution is non-equilibrium and back-action of the tunnel induced
photons on the transported electrons becomes important. This will
introduce new physical effects, beyond what was investigated in
earlier works where electronic transport through conductors in the
presence of a thermalized electromagnetic environment was at the
focus.
\cite{Nazarov92,Delsing89,Girvin90,Devoret90,Cleland90,Holst94,Basset10,Pashkin11,hofheinz11}
\\
\\
The \textit{ultrastrong coupling} regime in transport corresponds to a
coupling strength between the transport electrons and cavity photons
of the order of the frequency of the fundamental mode of the
cavity. In this regime electrons entering the conductor strongly
modify the photon states of the cavity and microwave polarons are
formed. To the best of our knowledge the ultrastrong coupling regime
has not been reached experimentally in conductor-cavity systems. In
this context it is interesting to point out the strong similarities
between the physics of transport through conductors coupled to
microwave cavities and molecular electronics and nano-electro
mechanics, where the conduction electrons couple to vibrational
degrees of freedom, or
phonons.\cite{Park00,Boese01,Braig03,Mitra04,Sapmaz06,Leturcq09} In fact, in
these type of systems ultrastrong electron-phonon coupling has
recently been demonstrated.\cite{Sapmaz06,Leturcq09} Several
non-trivial transport properties resulting from a non-equilibrium
phonon distribution has further been investigated theoretically in
this regime, e.g. super-poissonian \cite{Koch05} or suppressed \cite{Haupt06} shot-noise and
negative differential
conductance. \cite{Boese01,Koch052,Koch06,Shen07} Moreover, the
non-equilibrium phonon distribution itself has been found to possess
non-trivial properties.\cite{Koch06,merlo08,ioffe08,hartle11,Piovano11} These
results clearly promotes investigations of electron-photon analogs of
electron-phonon phenomena, performed in strongly coupled
conductor-cavity systems.
\\
\\
Taken together these observations provide strong motivation for a
careful theoretical investigation of the regimes of \textit{strong} and
\textit{ultrastrong coupling} between electrical conductors and microwave cavities. 
In this work we present a detailed investigation of a conductor capacitively coupled to a
microwave cavity, focusing on the properties of the electronic transport
through the conductor and the transport-induced photon state in the cavity. 
The conductor is taken to be an electrostatically gated metallic dot,
a single electron transistor, in the normal state. The combined
all-metal dot-cavity system can be realized with existing lithographic
techniques, giving large experimental versatility when trying to
increase the coupling strength. Moreover, as we demonstrate in this
work, the metallic dot-cavity system allows for a detailed and
consistent strong coupling analysis, analytical as well as numerical,
of the deep quantum, few photon regime where interesting, new physical
phenomena are most clearly manifested. We point out that albeit
focusing on a metallic dot conductor, our approach can directly be
applied to few-level quantum dots.
\\
\\
In the first part of the paper we provide a detailed description of
the dot-cavity system and describe how to derive, based on the Lagrangian
formulation of circuit QED, a Hamiltonian for the isolated
dot-cavity system for arbitrary strong coupling. We demonstrate the
importance of a consistent strong-coupling treatment in order to avoid
unphysical effects that would follow from a naive extension of the
weak-coupling model to stronger couplings. We also discuss
possible experimental realizations of the strong
capacitive-coupling regime relevant for our model. For the dot coupled to
external leads, the total system is described by a quantum master equation
which accounts for both the electronic transport in the sequential tunneling
regime as well as the coherent, non-equilibrium dynamics of the photon state. We first
analyze the properties of the photon state for the cases where one and
two photon modes in the cavity are active. For a single active mode we
describe the transport-induced photon state for different dot-cavity
couplings, focusing on the non-equilibrium distribution and the
signatures of microwave polaron formation. Analytical results are
obtained in the limit where the coupling strength is small 
compared to the fundamental frequency of the cavity. For two active 
modes we investigate inter-mode conversion of
photons and in particular the coherence properties of the photon
state, important in the ultrastrong coupling regime. An effective model for the
maximally coherent situation is
presented, allowing us to find accurate expressions for the photon
state also at ultrastrong couplings. Turning to the electron transport, the
conductance and the noise through the dot is analyzed for different
dot-cavity coupling strengths. For coupling strengths much
smaller than the fundamental frequency of the cavity the 
current and noise are shown to be independent on the photon state. For
stronger couplings the current and noise are compared to results for
an equilibrated photon state and we identify clear effects on the
transport due to the non-equilibrium photon state. Most prominently we
find super-poissonian noise at ultrastrong couplings, an indication of
the avalanche effect discussed for molecular electronics in
Ref. \onlinecite{Koch05}.

\section{System and method}
\begin{figure}[h]
\begin{center}
\includegraphics[trim=1cm 1.90cm 1.6cm 0cm, clip=true, width=0.45\textwidth, angle=0]{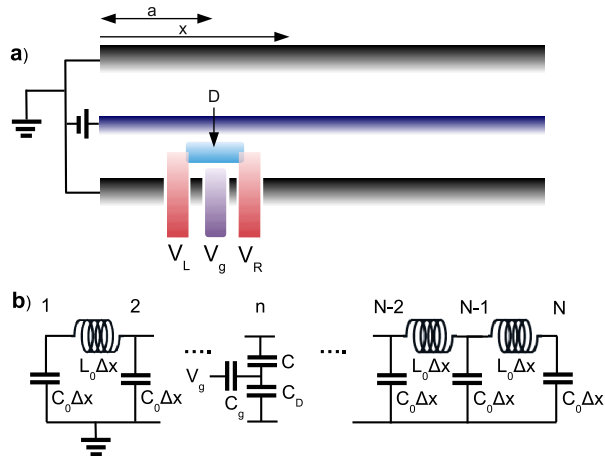}
\end{center}
\caption{(Color online) a) Schematic of the system with a normal state
  metallic dot (D) inserted in a transmission line cavity a distance
  $a$ from the left end. The central conductor (blue) can be a
  superconductor or e.g. an array of SQUIDS. The dot is tunnel coupled
  to two metallic leads L,R (red) kept at voltages $V_{L}$ and
  $V_{R}$, respectively. The dot is capacitively coupled to a gate
  electrode (purple) kept at the voltage $V_{g}$. b) Discrete circuit
  representation of the dot-cavity system with $N$ nodes. The dot is
  connected to node $n$. The inductance between two adjacent nodes is
  $L_{0}\Delta x$ and the capacitance to ground from node $i\neq n$ is
  $C_{0}\Delta x$, giving the cavity a characteristic impedance
  $\sqrt{L_{0}/C_{0}}$. The dot-cavity capacitance is $C$, the
  dot-to-ground capacitances is $C_{D}$ and the gate capacitance is
  $C_{g}$.}
\label{figsys}
\end{figure}

We consider the system shown in Fig. \ref{figsys}. A normal state
metallic dot is inserted between the central conductor and one of the
ground planes in a superconducting transmission line cavity. The
cavity has a length $d$ and the dot is placed a distance $a$ from the
left end. $C$ and $C_{D}$ denotes the capacitance between the dot and
ground, and between the dot and the cavity central conductor,
respectively. The cavity has a characteristic impedance $Z_{0}
=\sqrt{L_0/C_0}$, where $L_{0}$ and $C_{0}$ are the inductance and
capacitance per unit length. The central conductor can be made of a
superconducting material or e.g. a metamaterial, as a
SQUID-array.\cite{Castellanos07,Castellanos08} The dot is further
tunnel coupled to electronic leads $\ell=L,R$, kept at bias voltages
$V_{\ell}$. We assume that the lead-dot resistances are much larger
than the quantum resistance quantum $R_{q}=h/e^{2}$; the transport is
in the Coulomb blockade regime with a well defined charge on the dot.
The leads are assumed to be in thermal equilibrium at a temperature
$T$.  Moreover the electron relaxation rate of the dot is assumed to
be much shorter than the tunneling rate, i.e. the electrons reach
thermal equilibrium, at temperature $T$, in between each tunneling
event.  The background charge on the dot can be controlled with a gate
electrode, kept at a bias $V_{g}$, via a gate capacitance denoted
$C_{g}$.  The relaxation of the photons in the cavity due to electron
tunneling is much faster than the the intrinsic relaxation rate,
$\kappa$, in high quality cavities and we thus neglect all intrinsic
sources of photon loss.

\subsection{Cavity-dot system}

Our initial aim is to arrive at a Hamiltonian for the total system,
without any approximation on the dot-cavity coupling strength. We
start by considering the isolated dot-cavity system and derive a
Hamiltonian expressed in terms of the charge on the dot, the photons
in the cavity and the interaction between them. Following standard
circuit QED procedure \cite{Yurke,Devoret} we first write down the
Lagrangian for the circuit. We note that similar systems, with the
focus on arbitrary strong dot-cavity coupling have been treated in
e.g. Refs. \onlinecite{Falci,Lehur}. The discussion here is therefore
kept short and details are presented only where our derivation differs
from previous works.
\\
\\
The transmission line cavity is represented by a chain of $N\gg 1$
identical LC-circuits with capacitance $C_{0}\Delta x$ and inductance
$L_{0}\Delta x$, where $\Delta x=d/N$. The quantum dot is coupled to
the chain node $n=Na/d$ and to ground via capacitances $C$ and
$C_{D}$, respectively. The Lagrangian of the circuit is then
\begin{align}
&L=\sum_{i\neq n}^{N}\frac{C_{0}\Delta x\dot{\phi}_{i}^{2}}{2}-\sum_{i}^{N-1}\frac{(\phi_{i+1}-\phi_{i})^{2}}{2L_{0}\Delta
  x}+\frac{C_{D}\dot{\phi}_{D}^{2}}{2}+\nonumber\\
&+\frac{C(\dot{\phi}_{n}-\dot{\phi}_{D})^{2}}{2}+\frac{C_{g}(V_{g}+\dot{\phi}_{D})^{2}}{2}
\end{align}
where $\phi_{i}$ is the phase of the i:th node and $\phi_{D}$ the phase of the
dot. 
\\
\\
To find the normal modes\cite{Goldstein} of the combined cavity-dot system we
consider the Euler-Lagrange equations $\frac{d}{dt}\frac{\partial L}{\partial\dot{\phi}_{i}}-\frac{\partial
  L}{\partial\phi_{i}}=0$, for $i=1,..N,D$. Using the equation of $i=D$,
$\ddot{\phi}_{D}$ can be expressed in terms of $\ddot{\phi}_{N}$ and
substituted into the equation $i=N$. We can then write the equations for the cavity
phases in matrix form  
\begin{equation}
\mathcal{T}\ddphiv=\cal{V}\phiv
\label{mEuler}
\end{equation}
where $\phiv=[\phi_{1} .. \phi_{N}]^{T}$ and the matrices
$\cal{T},\cal{V}$ have elements
$\mathcal{T}_{ij}=\delta_{ij}\left[C_{0}\Delta
  x+\frac{C(C_{0}\Delta x+C_{g})}{C+C_{D}+C_{g}}\delta_{in}\right]$ and
$\mathcal{V}_{ij}=\frac{1}{L_{0}\Delta
  x}(\delta_{ij}(2-\delta_{i1}-\delta_{iN})-\delta_{i(j-1)}-\delta_{i(j+1)})$. Since
$\mathcal{T}$ is diagonal with positive elements and $\cal{V}$ is real
and symmetric we can express Eq. \eqref{mEuler} in the basis of normal
modes as $\ddvarphiv=\Lambda\varphiv$, where $\phiv=M\varphiv$. The
elements $\Lambda_{p}$ of the diagonal matrix $\Lambda$ are the
frequencies of the normal modes squared,
i.e. $\Lambda_{p}=\omega_{p}^{2}$. The columns, $\bf{m}_{p}$, in $M$
are the solutions to the eigenvalue problem
\begin{equation}
\mathcal{T}^{-1}\mathcal{V}{\bf{m}}_{p}=\omega_{p}^{2}{\bf{m}}_{p},
\label{geneig}
\end{equation}
with the normalization condition
${\bf{m}}_{p}^{T}\mathcal{T}{\bf{m}}_{q}=C_{0}d\delta_{pq}$. We can then express the
Lagrangian in terms of the normal modes as
\begin{align}
&L=\sum_{p}\left(\frac{C_{0}d\dot{\varphi}_{p}^{2}}{2}-\frac{C_{0}d\omega_{p}^{2}\varphi_{p}^{2}}{2}\right)+\frac{C^{2}}{2C_{\Sigma}}\sum_{pq}M_{np}M_{nq}\dot{\varphi}_{p}\dot{\varphi}_{q}\nonumber\\
&+C_{g}V_{g}\dot{\varphi}_{D}+\frac{C_{D}\dot{\varphi}_{D}^{2}}{2} -C\dot{\varphi}_{D}\sum_{p}M_{np}\dot{\varphi}_{p}
\end{align}
where $C_{\Sigma}=C_{D}+C+C_{g}$ and we write $\phi_{D}=\varphi_{D}$
for notational convenience.
\\
\\
In the continuum limit, $N\rightarrow\infty$, $\Delta x\rightarrow 0$
with $N\Delta x=d$ constant, the vectors $\bf{m}_{p}$ turn into
continuous functions, $\zeta_{p}(x)$, of the coordinate $x$ along the
transmission line.  From Eq.~\eqref{geneig} it is found that the
functions $\zeta_{p}(x)$ satisfy the differential equation
\begin{align}
&\zeta_{p}''(x)+k_{p}^{2}[1+d\alpha\delta(x-a)]\zeta_{p}(x)=0,
\label{wave}
\end{align}
with boundary conditions $\zeta_{p}^{'}(0)=\zeta_{p}^{'}(d)=0$.
Here $k_{p}=\sqrt{L_{0}C_{0}}\omega_{p}$ and
$\alpha=C_{g}C/(C_{\Sigma}C_{0}d)$. The normalization
condition above becomes   
\begin{equation}
\frac{1}{d}\int_{0}^{d}dx\zeta_{p}(x)\zeta_{q}(x)[1+\alpha
d\delta(x-a)]=\delta_{pq}.
\label{normalization}
\end{equation}
This generalized Sturm-Liouville problem has solutions
\begin{align}
\zeta_{p}(x)=\left\{\begin{array}{cc}
    A_{p}\cos(k_{p}x) & 0\leq x\leq a\\
B_{p}\cos[k_{p}(d-x)] & a\leq x\leq d,
\end{array}\right.
\label{contmode}
\end{align}
where $A_{p}\cos(k_{p}a)=B_{p}\cos[k_{p}(d-a)]$ and $k_{p}$ are the positive
solutions to the equation
\begin{equation}
\frac{\tan(k_{p}d)[1+\tan^2(k_{p}a)]}{1+\tan(k_{p}a)\tan(k_{p}d)}=-\alpha k_{p}d,
\label{transeq}
\end{equation}
following from Eq.~\eqref{wave} with boundary conditions. The
solutions are illustrated in Fig. \ref{renorm}. The normalization
condition in Eq.~\eqref{normalization} gives
\begin{equation}
A_{p}^{2}=\frac{d}{2}\left[a+\frac{\sin(2k_{p}a)}{2k_{p}}+\cos^{2}(k_{p}a)F_{p}\right]^{-1}
\label{norm}
\end{equation}
with
\begin{equation}
F_{p}=\frac{(d-a)}{\cos^{2}[k_{p}(d-a)]}+\frac{\tan[k_{p}\left(d-a\right)]}{2k_{p}}+\alpha d.
\end{equation}
We see that in the limit $\alpha k_{p}d\ll 1$, corresponding to low
frequencies $\omega_{p}$,  the solutions $k_{p}$ in Eq.~\eqref{transeq}
approach $p\pi/d$, the result for the cavity disconnected from the dot.
In the opposite limit, $\alpha k_{p}d\gg 1$, the solutions approach $(p+1/2)\pi/a$
and $(p+1/2)\pi/(d-a)$. This gives, from  Eq.~\eqref{contmode} that the
amplitudes $\zeta_{p}(x)$, at $x=a$, will be zero: for large $\omega_{p}$ the cavity is
effectively grounded via the dot. 
\\
\\
Thus, in the continuum limit we obtain the Lagrangian for the system
\begin{align}
&L=\sum_{p}\left(\frac{C_{C}\dot{\varphi}_{p}^{2}}{2}-\frac{(k_{p}d)^{2}\varphi_{p}^{2}}{2L_{C}}\right)-C\dot{\varphi}_{D}\sum_{p}\zeta_{p}(a)\dot{\varphi}_{p}\nonumber\\
&+\frac{C^{2}}{2C_{\Sigma}}\sum_{pq}\zeta_{p}(a)\zeta_{q}(a)\dot{\varphi}_{p}\dot{\varphi}_{q}+\frac{C_{\Sigma}\dot{\varphi}_{D}^{2}}{2}+C_{g}V_{g}\dot{\varphi}_{D},\nonumber\\
\end{align}
where $C_{C}=C_{0}d$ and $L_{C}=L_{0}d$ are the total capacitance and
inductance of the cavity. This Lagrangian can now be used to obtain
the conjugate variables $Q_{D}=\partial L /\partial \dot{\varphi}_{D}$
and $Q_{p}=\partial L/\partial \dot{\varphi}_{p}$ to $\varphi_{D}$ and
$\varphi_{p}$ respectively.  We point out that $Q_{D}$ is the charge
on the dot.

Expressing $\dot{\varphi}_{D}$ and $\dot{\varphi}_{p}$ in terms of $Q_{D}$ and
$Q_{p}$ and using the Legendre transformation,
$H_{S}=Q_{D}\dot{\varphi}_{D}+\sum_{p}Q_{p}\dot\varphi_{p}-L$, the
following classical Hamiltonian of the system is obtained:
\begin{align}
&H_{S}=\sum_{p}\left(\frac{Q_{p}^{2}}{2C_{C}}+\frac{(k_{p}d)^{2}\varphi_{p}^{2}}{2L_{C}}+\frac{C(Q_{D}-C_{g}V_{g})}{C_{\Sigma}C_{0}}\zeta_{p}(a)Q_{p}\right)\nonumber\\
&+\frac{(Q_{D}-C_{g}V_{g})^{2}}{2C_{\Sigma}}\left(1+\frac{C^{2}}{C_{\Sigma}C_{C}}\sum_{p}\zeta_{p}(a)^{2}\right).
\end{align}
The quantum Hamiltonian is obtained by canonical quantization. The
generalized coordinates $Q_{p},\varphi_{p},Q_{D}$ and $\varphi_{D}$ are replaced by 
operators $\hat{Q}_{p},\hat{\varphi}_{p},\hat{Q}_{D}$ and $\hat{\varphi}_{D}$ and 
the commutation relations $[\hat{Q}_{p},\hat{\varphi}_{q}]=i\hbar\delta_{pq}$, for
$p,q=D,1,2,..$, are imposed. For the coordinates of the cavity
$\hat{\varphi}_{p},\hat{Q}_{p}$ creation and annihilation operators
$\hat{a}_{p},\hat{a}_{p}^{\dag}$ are introduced for $p=1,2,..$ according to
\begin{align}
&\hat{Q}_{p} =\sqrt{\hbar
  k_{p}d}\left(\frac{C_{0}}{L_{0}}\right)^{1/4}\frac{(\hat{a}_{p}^{\dag}+\hat{a}_{p})}{\sqrt{2}}\nonumber\\
&\hat{\varphi}_{p}=i\sqrt{\frac{\hbar}
  {k_{p}d}}\left(\frac{L_{0}}{C_{0}}\right)^{1/4}\frac{(\hat{a}_{p}^{\dag}-\hat{a}_{p})}{\sqrt{2}}.
\end{align} 
These operators fulfill bosonic commutation relations
$[\hat{a}_{p},\hat{a}^{\dag}_{q}]=\delta_{pq}$. The Hamiltonian of the isolated dot-cavity system can then be written
\begin{align}
&\hat{H}_{S}=\underbrace{\sum_{p}\hbar\omega_{p}\hat{a}_{p}^{\dag}\hat{a}_{p}}_{\hat{H}_{C}}+\underbrace{\frac{(\hat{Q}_{D}-C_{g}V_{g})^{2}}{2C_{\Sigma}}\left(1+\frac{C^{2}}{C_{\Sigma}C_{C}}\sum_{p}\zeta_{i}^{2}(a)\right)}_{\hat{H}_{D}}\nonumber\\
&+\underbrace{\frac{C(\hat{Q}_{D}-C_{g}V_{g})}{C_{\Sigma}}\sum_{p}\sqrt{\frac{\hbar\omega_{p}}{2C_{C}}}(\hat{a}_{p}+\hat{a}_{p}^{\dag})\zeta_{p}(a)}_{\hat{H}_{DC}}.
\label{Hc}
\end{align}
This Hamiltonian has the desired form
$\hat{H}_{S}=\hat{H}_{C}+\hat{H}_{D}+\hat{H}_{DC}$. The first term, $\hat{H}_{C}$, is the Hamiltonian of a set of harmonic oscillators corresponding to
cavity modes with frequencies $\omega_{p}$. These frequencies are obtained by solving Eq.
\eqref{transeq}. The second term, $\hat{H}_{D}$, corresponds to the charging
energy of the dot. We see that this is larger than for a dot with
self-capacitance $C_{\Sigma}$. The third term in the Hamiltonian,
$\hat{H}_{DC}$, is the linear coupling between the charge of the dot and the
modes in the cavity. It is convenient for the further analysis to introduce
the dimensionless coupling constant
\begin{equation}
\lambda_{p}=\frac{C}{C_{\Sigma}}\frac{e\zeta_{p}(a)}{\sqrt{2\hbar\omega_{p}C_{C}}}=\frac{C}{C_{\Sigma}}\sqrt{\frac{Z_{0}}{R_{q}}}\frac{\zeta_{p}(a)}{\sqrt{k_{p}d/(2\pi)}},
\label{lam}
\end{equation}
We emphasize that the Hamiltonian Eq.~\eqref{Hc} has been obtained in an exact way, without
any assumptions about the cavity-dot coupling strength.
It is interesting to note, just as was done in
Ref. \onlinecite{Falci}, that this exact treatment gives a
Caldeira-Leggett type Hamiltonian, naturally including
the so called counter term. \cite{Weissbook} This counter term is
typically introduced by hand to ensure a spatially uniform damping
in the Caldeira-Leggett model. In our model the counter term just
comes from the part of the charging energy term $\hat H_D$ arising
from the normalization of the capacitance $C_{\Sigma}$.

\subsection{Coupling to leads and Lang-Firsov transformation}

As a next step we consider the tunnel coupling of the dot to external
leads $L$ and $R$. Following the standard path for transport through
single-electron-transistors, \cite{Nazarov92} the orbital and charge degrees
of freedom of the metallic dot are treated separately. We describe the orbital
degrees of freedom by the Hamiltonian
\begin{equation}
H_{O}=\sum_{k'}\epsilon_{Dk'}\hat{c}^{\dag}_{Dk'}\hat{c}_{Dk'},
\label{H0} 
\end{equation}
where $\hat{c}_{Dk'}^{\dag}$ creates an electron with energy $\epsilon_{Dk'}$ in the dot. The 
Hamiltonian of the leads is
\begin{equation}
\hat{H}_{L}=\sum_{\ell,k}\epsilon_{\ell
  k}\hat{c}^{\dag}_{\ell k}\hat{c}_{\ell k},
\label{HL}
\end{equation}
where $\hat{c}^{\dag}_{\ell k}$ is the creation operator of an
(uncharged) electron with energy $\epsilon_{\ell k}$ in lead
$\ell=L,R$. In Eqs.~\eqref{H0} and~\eqref{HL} the indices $k$ and $k'$
denotes both wave number and spin. The tunnel Hamiltonian is written
as
\begin{equation}
\hat{H}_{T}=\sum_{\ell,k,k'}t_{\ell kk'}\hat{c}_{\ell k}^{\dag}\hat{c}_{Dk'}\exp\left(\frac{ie\hat{\varphi}_{D}}{\hbar}\right)+h.c,
\end{equation}   
where the operators $\exp\left(\mp ie\hat{\varphi}_{D}/\hbar\right)$
has the effect of changing the dot charge by $\pm 1$. This yields a
Hamiltonian of the total system
\begin{equation}
\hat{H}=\hat{H}_{O}+\hat{H}_{C}+\hat{H}_{D}+\hat{H}_{DC}+\hat{H}_{T}+\hat{H}_{L}.
\end{equation}
For further analysis it is convenient to first perform a canonical
transformation of $\hat{H}$ that removes the linear-in-charge term
$\hat{H}_{DC}$. Such a Lang-Firsov, or polaron,\cite{mahan}
transformation is carried out by transforming the Hamiltonian as
$\bar{H}=\exp(\hat{s})\hat{H}\exp(-\hat{s})$ and state kets as
$\ket{\bar{\Psi}}=\exp(-\hat{s})\ket{\Psi}$ with
$\hat{s}=[(\hat{Q}_{D}-C_{g}V_{g})/e]\sum_{p}\lambda_{p}(\hat{a}_{p}^{\dag}-\hat{a}_{p})$.
We then arrive at the Hamiltonian
\begin{align}
&\bar{H}=\hat{H}_{L}+\hat{H}_{O}+\sum_{p}\hbar\omega_{p}\hat{a}_{p}^{\dag}\hat{a}_{p}+\frac{(\hat{Q}_{D}-C_{g}V_{g})^{2}}{2C_{\Sigma}}\nonumber\\
&\!\!\!\!\!\!+\sum_{\ell,k,k'}t_{\ell kk'}\hat{c}_{\ell k}^{\dag}\hat{c}_{D
  k'}\exp\left(\frac{ie\hat{\varphi}_{D}}{\hbar}\right)\exp\left[-\sum_{p}\lambda_{p}(\hat{a}_{p}^{\dag}-\hat{a}_{p})\right]\nonumber\\
&+h.c,
\label{fullHam1}
\end{align}
The eigenstates of the isolated dot-cavity system, decoupled from the leads,
are up to an unimportant phase factor given by
\begin{equation}
\ket{N\bold{n}}=\ket{N}_{el}\exp\left[N\sum_{p}\lambda_{p}(\hat{a}_{p}^{\dag}-\hat{a}_{p})\right]\ket{\bold{n}},
\end{equation}
the tensor product of the charge state with $N$ excess electrons on
the dot, $\ket{N}_{el}$, and the Fock states of the cavity modes,
$\ket{\bold{n}}=\ket{n_{1}n_{2}...}$, displaced by $N\lambda_{p}$
each. We refer to the states $\ket{N\bold{n}}$ as microwave polaron
states and $n_{p}$ as the number of photons in mode $p$. The energies
of the polarons are given by
\begin{equation}
\epsilon_{N\bold{n}}=\frac{e^{2}(N-n_{g})^{2}}{2C_{\Sigma}}+\sum_{p}n_{p}\hbar\omega_{p}
\label{eigen}
\end{equation}
with $n_{g}=C_{g}V_{g}/e$.  Looking at Eq.~\eqref{eigen} we note that
the shift in charging energy from the coupling to the cavity modes, a
polaron shift, is exactly canceled by the extra charging energy due to
the renormalization of the capacitance of the dot. This cancellation
is a direct consequence of the exact treatment of the cavity-dot
coupling throughout the derivation. If one instead of Eq.~\eqref{Hc}
naively would start with a standard Anderson-Holstein type
Hamiltonian, i.e. without the renormalized capacitance $C_{\Sigma}$,
and then perform the polaron transformation, the resulting charging
energy term could become negative for large dot-cavity couplings. For
a metallic dot, with a continuous density of states, such a model
would be unphysical; the system would lack a well defined ground state
since increasing the number of electrons on the dot always would lower
the total energy of the system. It should be noted that problems with
infinite negative energies typically do not appear in related
electron-phonon models in molecular electronics.\cite{note1}

\section{Quantum master equation}

From the Hamiltonian in Eq. \eqref{fullHam1} we can then derive a quantum master
equation describing the dynamics of both the charge state in the dot and the 
state of the cavity modes. The derivation follows a standard path, see
e.g. Refs. \onlinecite{qnoise,Mitra04,Rodrigues05,Hubener09}. 

\subsection{Derivation}
In the rest of the paper we consider the case where the charging
energy of the dot, $e^{2}/(2C_{\Sigma})$, is the largest energy in the
system. It is then safe to assume that the number of excess electrons
on the dot will only fluctuate between $N$ and $N+1$. For simplicity,
we consider gate voltages such that $N$ can only take values 0 and
1. The difference in charging energy between states with $0$ and $1$
electrons is denoted $\Delta E_{C}$.
\\
\\
Starting from the Liouville equation for the density matrix, expanding
to leading order in tunnel-coupling and tracing over reservoir and
fermionic dot degrees of freedom we arrive at a quantum master
equation for the elements of the reduced density matrix $\rho$ of the
dot-cavity system. A more detailed derivation is presented in Appendix
A. This equation is in the polaron basis given by
\begin{align}
  &\frac{d}{dt}\braket{0\bold{n}|\rho
    |0\bold{m}}=-\frac{i}{\hbar}(\epsilon_{0\bold{n}}-\epsilon_{0\bold{m}})\braket{0\bold{n}|\rho
    |0\bold{m}}\nonumber\\
  +\sum_{\ell,\bold{k},\bold{l}}&\Gamma_{\ell}[-h_{\ell}(\epsilon_{1\bold{l}}-\epsilon_{0\bold{k}})\prod_{p}X^{p}_{k_{p}l_{p}}X^{p}_{m_{p}l_{p}}\braket{0\bold{n}|\rho |0\bold{k}}\nonumber\\
  &+g_{\ell}(\epsilon_{1\bold{l}}-\epsilon_{0\bold{m}})\prod_{p}X^{p}_{n_{p}k_{p}}X^{p}_{m_{p}l_{p}}\braket{1\bold{k}|\rho |1\bold{l}}\nonumber\\
  &+g_{\ell}(\epsilon_{1\bold{k}}-\epsilon_{0\bold{n}})\prod_{p}X^{p}_{n_{p}k_{p}}X^{p}_{m_{p}l_{p}}\braket{1\bold{k}|\rho |1\bold{l}}\nonumber\\
  &-h_{\ell}(\epsilon_{1\bold{k}}-\epsilon_{0\bold{l}})\prod_{p}X^{p}_{n_{p}k_{p}}X^{p}_{l_{p}k_{p}}\braket{0\bold{l}|\rho|0\bold{m}}]\nonumber
\end{align}
\begin{align}
&\frac{d}{dt}\braket{1\bold{n}|\rho|1\bold{m}}=-\frac{i}{\hbar}(\epsilon_{1\bold{n}}-\epsilon_{1\bold{m}})\braket{1\bold{n}|\rho
  |1\bold{m}}\nonumber\\
+\sum_{\ell,\bold{k},\bold{l}}&\Gamma_{\ell}[-g_{\ell}(\epsilon_{1\bold{k}}-\epsilon_{0\bold{l}})\prod_{p}X^{p}_{l_{p}k_{p}}X^{p}_{l_{p}m_{p}}\braket{1\bold{n}|\rho |1\bold{k}}\nonumber\\
&+h_{\ell}(\epsilon_{1\bold{m}}-\epsilon_{0\bold{l}})\prod_{p}X^{p}_{k_{p}n_{p}}X^{p}_{l_{p}m_{p}}\braket{0\bold{k}|\rho |0\bold{l}}\nonumber\\
&+h_{\ell}(\epsilon_{1\bold{n}}-\epsilon_{0\bold{k}})\prod_{p}X^{p}_{k_{p}n_{p}}X^{p}_{l_{p}m_{p}}\braket{0\bold{k}|\rho |0\bold{l}}\nonumber\\
&-g_{\ell}(\epsilon_{1\bold{l}}-\epsilon_{0\bold{k}})\prod_{p}X^{p}_{k_{p}n_{p}}X^{p}_{k_{p}l_{p}}\braket{1\bold{l}|\rho |1\bold{m}}]\nonumber\\
\label{gme}
\end{align}
where $h_{\ell}(x)=(x-[\mu_{\ell}-\mu_{D}])/(\hbar\omega_{1})[\exp[(x-[\mu_{\ell}-\mu_{D}])/k_{B}T]-1]^{-1}$,
$g_{\ell}(x)=\exp[(x-[\mu_{\ell}-\mu_{D}])/k_{B}T]h_{\ell}(x)$ and
$\Gamma_{\ell}=2\pi|t_{\ell}|^{2}\nu_{\ell}\nu_{D}\omega_{1}$. Here,
$\mu_{\ell}$ and $\mu_{D}$ are the chemical potentials of the leads and the
dot, respectively. Moreover, we have assumed tunneling amplitudes independent of lead and dot
energy, i.e.  $t_{kk'\ell}\approx t_{\ell}$ and $\nu_{\ell}$,
$\nu_{D}$ denotes the density of states of lead $\ell$  and the dot, respectively. Furthermore
\begin{align}
&X^{p}_{nm}=\braket{n|\exp[-\lambda_{p}(\hat{a}^{\dag}_{p}-\hat{a}_{p})]|m}=\frac{1}{\sqrt{m!}}e^{-\lambda_{p}^{2}/2}\nonumber\\
&\times\sum_{j=0}^{min(m,n)}\lambda^{n+m-2j}_{p}(-1)^{n-j}\bin{m}{j}\frac{\sqrt{n!}}{(n-j)!}
\label{Franck}
\end{align}
are the Franck-Condon factors\cite{mahan} for the p:th mode. These are
the amplitudes for the transition from the state in mode $p$ going
between polaron states with $n$ and $m$ quanta as the electron tunnels
into or our of the dot. Formally $X^{p}_{nm}$ is given by the overlap
of oscillator wavefunctions before and after the tunneling. We
emphasize that Eq.~\eqref{gme} is a quantum master equation: it
describes the dynamics of the polaron states as well as coherences
between them.

\subsection{Franck-Condon effect}

From Eq.~\eqref{Franck} we note that for all Franck-Condon factors
$X^{p}_{nm}\propto\exp(-\lambda_{p}^{2}/2)$.  This means that even if
no photons are excited as the electrons tunnel into and out of the
dot, $n=m=0$, the presence of the modes in the cavity will still
affect transport via renormalized, suppressed tunneling rates. This
Franck-Condon suppression of electron tunneling is a pure vacuum
effect, a consequence of the tunneling charge having to displace all
the oscillators in the cavity. We introduce the vacuum renormalized
tunneling rates
\begin{equation}
\tilde{\Gamma}_{\ell}=\Theta\Gamma_{\ell},\quad\Theta=\exp(-\sum_{p}\lambda_{p}^2),
\label{renormeq}
\end{equation}
where $\Theta$ denotes the renormalization factor.
It is convenient to also introduce the notation
$Y^{p}_{nm}=\exp(\lambda_{p}^{2}/2)X^{p}_{nm}$ for the remaining part of the
Franck-Condon factors for the $p$:th mode.
\\
\begin{figure}
\centering
\includegraphics[trim=0cm 0cm -1.5cm 0cm, clip=true, width=0.45\textwidth, angle=0]{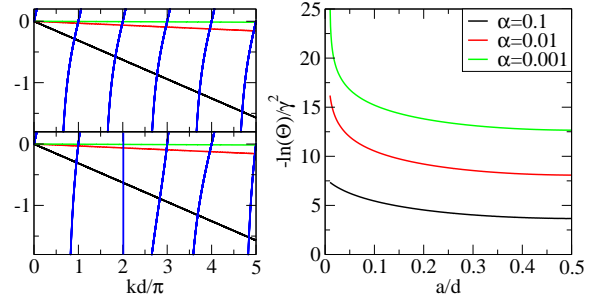}
\caption{(Color online) Left panels: Intersections of the left and
  right hand side of Eq. \eqref{transeq}, for different $\alpha$ (see
  right panel inset). The ratios $a/d=0$ and $a/d=1/4$ for upper and
  lower panel respectively. Right panel: Position dependence of the
  renormalization factor $-\ln(\Theta)/\gamma^{2}$, with
  $\gamma=(C/C_{\Sigma})\sqrt{Z_{0}/R_{q}}$. Only positions $0\leq
  a\leq d/2$ are plotted since the renormalization factor is symmetric
  with respect to $d/2$.}
\label{renorm}
\end{figure}
\\
From Eq.~\eqref{lam} it follows that the coupling constant,
$\lambda_{p}$, is proportional
$\zeta_{p}(a)/\sqrt{k_{p}d}$. Consequently [see Eqs.~\eqref{contmode}
and~\eqref{transeq}], the renormalization factor depends on the the
distance $a$ and the parameter $\alpha=CC_{g}/(C_{\Sigma}C_{C})$.
Since the dot can be placed at any position $a$, or effectively be
moved by tuning the boundary conditions of the
cavity,\cite{Wallquist06,Sandberg08} it is interesting to study the
position dependence of $\Theta$, plotted in Fig.~\ref{renorm} for
different values of $\alpha$. Several observations can be made: $(i)$
Albeit the renormalization factor can be large, it is always finite
for $\alpha >0$. There is thus no tunneling orthogonality catastrophe,
i.e zero over-lap between initial and final state in a tunneling
event. Such an orthogonality catastrophe would occur if one naively
replaces $\zeta_{p}(a)$ and $k_{p}$ with the corresponding amplitude,
$\sqrt{2/d}\cos(p\pi a/d)$, and wavenumber $p\pi/d$, of the cavity
disconnected from the dot. The exponent of the renormalization factor
would then be proportional to $\sum_{p}\cos(p\pi/d)^2/p$ which
diverges logarithmically. We emphasize that it is our exact treatment
of the dot charge-cavity coupling fully taking into account the effect
of the presence of the dot on the cavity modes that gives a finite
$\tilde{\Gamma}_{\ell}$.  $(ii)$ We see that the renormalization
factor has a strong dependence on the distance $a$, with a minimum at
$a=d/2$ and maximum at $a=0$. This is a consequence of that all modes
have maximal amplitude, $\zeta_{p}(a)$, at $ a=d$, while at $a=d/2$
half of the modes, i.e. the anti-symmetric, will have zero
amplitude. $(iii)$ We note that $\Theta$ decreases with decreasing
$\alpha$. This is to be expected, since a small $\alpha$ means that
the amplitude of the cavity modes at the connection point remains
finite for higher frequencies. It is also interesting to point out
that a position dependence of the coupling constant was very recently
investigated in the context of nano-electro mechanical
systems. \cite{Sasseti11,Donarini}

\subsection{Parameter regime}

The quantum master equation in Eq.~\eqref{gme} allows us to
investigate the charge and photon dynamics in a broad range of
parameters. The main interest of the present work is to investigate
new physical phenomena becoming important for strong dot-cavity
coupling.  This motivates us to focus on the deep quantum regime, with
only a few photons in the cavity, where these phenomena can be
investigated both qualitatively and quantitatively. Spelling out
explicitly the parameter range, we consider symmetric tunnel
couplings, $\Gamma_L=\Gamma_R=\Gamma$, and a symmetric bias
$\mu_L=-\mu_R=eV/2$, giving a chemical potential of the dot
$\mu_D=0$. We also consider the case where only the two lowest photon
modes have finite populations. This restriction puts limits on the
bias voltage; a careful investigation gives that $|eV/2\pm \Delta
E_{C}|<\hbar\omega_{2}$ is necessary to guarantee a negligible
occupation of the third and higher modes in all cases of interest.
This condition means that it is energetically forbidden for a
tunneling electron to emit a photon directly into the second
mode. However, population of the second mode is still possible by
inter-mode conversion of photons from the first mode, as discussed
below.  In the rest of the article we will use the simplified notation
$\ket{N n_{1} n_{2}}$ for the polaron states with $N=0,1$ electrons
and $n_{1}, n_{2}$ photons in the first and second mode, respectively.
\\
\\
We further assume that the tunneling rate is much smaller than the
fundamental cavity frequency, i.e $\tilde{\Gamma}\ll\omega_{1}$. For
the case where only the first photon mode is active, the off-diagonal
elements $\braket{\mu n_{1} 0|\rho|\mu m_{1}0}$, with $n_1\neq m_1$,
of the steady state density matrix in Eq.~\eqref{gme} are a factor
$\sim \tilde \Gamma/\omega_1\ll 1$ smaller than the diagonal elements
and can be disregarded.  This amounts to performing a secular, or
rotating-wave, approximation and reduces Eq.~\eqref{gme} to a standard
master equation.  For two active modes the situation is different
since two polaron states $\ket{N n_{1} n_{2}}$ and $\ket{N
  m_{1}m_{2}}$ can be degenerate, i.e. for
$n_{1}\omega_{1}+n_{2}\omega_{2}-(m_{1}\omega_{1}+m_{2}\omega_{2}) \ll
\tilde \Gamma $ the secular approximation can not be performed. The
off-diagonal density matrix elements $\braket{N n_{1} n_{2}|\rho|N
  m_{1}m_{2}}$, corresponding to coherences between polaron states
with different number of photons, must thus be retained in
Eq.~\eqref{gme}.  The simplest case giving degeneracy, discussed in
detail below, occurs for $\alpha \ll 1$ when from Eq.~\eqref{transeq}
$\omega_{2} \approx 2\omega_{1}$. Moreover, to highlight the effect of
the coherences we compare in several cases below the results based on
Eq.~\eqref{gme} to the results based on a master equation where the
off-diagonal elements are disregarded from the outset.
\\
\\
A key parameter in our work is the coupling constant, $\lambda_1$. To
reach the strong coupling regime, the time scale for tunnel induced
excitation and relaxation of the cavity photons must be much shorter
than the intrinsic relaxation time. This amounts to the restriction
\begin{equation}
\sqrt{\kappa/\tilde{\Gamma}}\ll\lambda_{1},
\label{ineqstrong}
\end{equation}
on the coupling constant, where $\kappa=\omega_{1}/(2\pi Q)$ is the
intrinsic relaxation rate of the first cavity mode and $Q$ the quality
factor. To provide a concrete estimate, for reasonable parameters of a
superconducting transmission line cavity $\omega_{1}/2\pi=10$GHz,
$Q=10^{6}$, $Z_{0}=100\Omega$ and $C\sim C_{\Sigma}$ one has
$\kappa=10$kHz and $\lambda_1=0.06$. Then, for a tunneling rate
$\tilde{\Gamma}=200$MHz, the left-hand side of Eq.~\eqref{ineqstrong}
is an order of magnitude smaller than the right-hand side.
\\
\\
The ultrastrong regime requires the coupling constant $\lambda_{1}$ to
be of order unity. For the capacitive dot-cavity coupling considered
here it has however been pointed out \cite{devoret07,Schoel2008} that
standard superconducting transmission lines only allow couplings
$\lambda_1$ up to a few percent. The limiting factor, clear from
Eq.~\eqref{lam}, is the ratio $Z_0/R_q \ll 1$.  To reach larger
couplings one thus has to consider ways of increasing the
characteristic impedance $Z_0$ of the transmission line.  One
promising possibility is transmission lines with a central conductor
consisting of an array of Josephson junctions or SQUIDS acting as
linear inductors. In recent experiments with SQUID array conductors
\cite{Castellanos07,Castellanos08} $Z_0 \approx 6k\Omega$,
i.e. $Z_0/R_q \approx 0.25$ was demonstrated, which would correspond
to $\lambda_1$ of the order of tens of percent for a dot capacitively
coupled to the transmission line. It should however be pointed out
that in such high impedance transmission lines non-linear effects, not
accounted for in our model, start to become relevant.
\\
\\
The relation between the coupling constants $\lambda_{1}$ and
$\lambda_{2}$ is determined by Eq.~\eqref{lam} and Eq.~\eqref{transeq}
as
\begin{equation}
\frac{\lambda_{2}}{\lambda_{1}}\approx\frac{\cos(2\pi
  a/d)}{\cos(\pi a/d)},
\label{coupl}
\end{equation}
for $\alpha\ll 1$. This relation is thus specified by $a$. Below we will
consider two important qualitatively distinct cases, $a=d/4$ and and $a=0$. For
$a=d/4$ we have $\lambda_{2}=0$ and only a the first mode has 
finite population. The case $a=0$
corresponds to a position in the cavity yielding maximal coupling strength. Eq.~\eqref{coupl} then gives $\lambda_{1}=\sqrt{2}\lambda_{2}$ and both the first and the 
second mode can have finite population.

\section{State of the photon modes of the cavity}

We first consider the current-induced photon state in the cavity, the
electronic transport is considered below. Experimentally, the photon
state in the cavity can e.g. be investigated by capacitively coupling
the cavity to a transmission line and measuring the state of the
output itinerant modes.\cite{bozyigit10} This gives access to the
frequency resolved population,\cite{hofheinz11} as well as higher
moments of the cavity field via e.g. quantum state tomography of
one\cite{Eichler10} or two\cite{Eichler11} itinerant modes. Moreover,
the photon number \cite{hofheinz08} as well as the full photon state,
\cite{hofheinz09} can also be obtained by coupling the cavity to a
superconducting qubit, embedded in the cavity. Studying specific
experimental setups to extract information about the photon state is
however out of the scope of the present article.  Hence we concentrate
on the photon state of the cavity described by the steady-state
density matrix, obtained from Eq.~\eqref{gme}.

\subsection{Single-mode}

\begin{figure}[h]
\includegraphics[trim=0cm 0cm 0cm 0cm, clip=true, width=0.45\textwidth, angle=0]{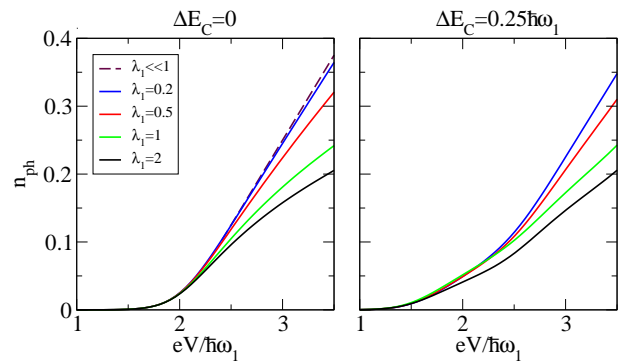}
\caption{(Color online) Mean number of photon excitations $n_{ph}$ for a single active mode
  $(\lambda_{2}=0)$ as a function of bias voltage $eV$ for different coupling
  strengths $\lambda_{1}$ and charging energy differences $\Delta
  E_{C}=0$ (left) and $0.25\hbar\omega_{1}$ (right). The
  temperature is $k_{B}T=0.05\hbar\omega_{1}$. In the left panel the dashed
  line gives the analytical result Eq. \eqref{anamean} for $eV-2\hbar\omega_{1}\gg k_{B}T$.}
\label{meanphoton1}
\end{figure}

\begin{figure}[h]
\includegraphics[trim=0cm 0cm 0cm 0cm, clip=true, width=0.45\textwidth, angle=0]{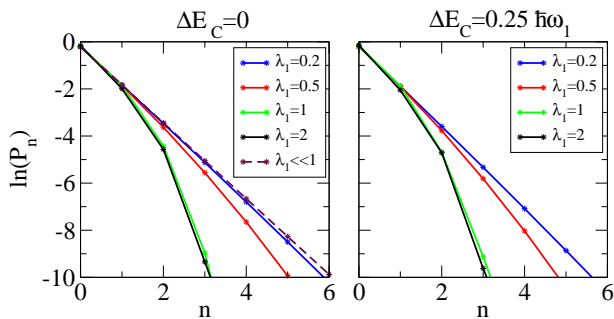}
\caption{(Color online) Logarithm of the probability of $n$ photons for different coupling
  strengths and $\Delta E_{C}=0$ (left) $\Delta E_{C}=0.25\hbar\omega_{1}$ (right) for bias voltage $eV=3\hbar\omega_{1}$ and $k_{B}T=0.05\hbar\omega_{1}$.}
\label{prob}
\end{figure}

\begin{figure}[h]
\includegraphics[trim=0cm 0cm 0cm 0cm, clip=true, width=0.45\textwidth, angle=0]{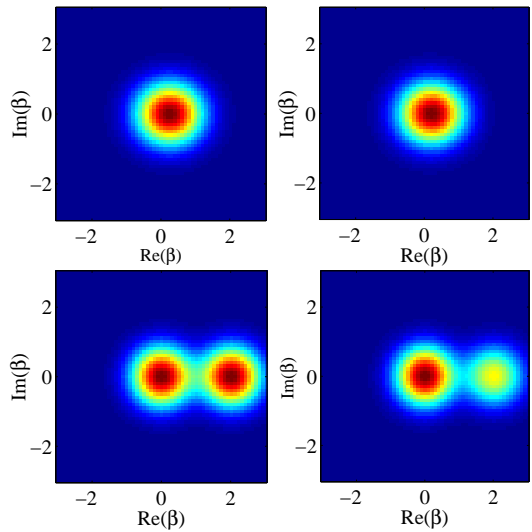}
\caption{(Color online) Wigner function $W(\beta)$ for coupling
  strengths $\lambda_{1}=0.2$ (upper panel) and $2$ (lower panel) and
  charging energy differences $\Delta E_{C}=0$ (left) and
  $0.25\hbar\omega_{1}$ (right). The color scale goes from blue (small
  value) to red (large value). The bias voltage is
  $eV=3\hbar\omega_{1}$ and the temperature is
  $k_{B}T=0.05\hbar\omega_{1}$.}
\label{Q}
\end{figure}

We first consider the case of a single active mode, obtained when the
coupling strength for the second mode is zero, i.e $\lambda_{2}=0$. To
demonstrate the effect of the tunneling electrons on the state of the
first mode it is instructive to consider the average number of photon
excitations in the two polaron states, $n_{ph}=\sum_{n}nP_{n}$ with
$P_{n}=\braket{0 n 0|\rho_{s}|0 n 0}+ \braket{1 n 0|\rho_{s}|1 n 0}$
and $\rho_{s}$ the steady-state density matrix. The average number of
excitations, $n_{ph}$, is related to the photon population in the
unrotated basis $\braket{\hat{n}_{1}}$ as
$\braket{\hat{n}_{1}}=n_{ph}+\lambda_{1}^{2}\sum_{n}\braket{1 n
  0|\rho_{s}|1 n 0}$. In Fig.~\ref{meanphoton1} $n_{ph}$ is plotted
against the bias voltage for different coupling strengths,
$\lambda_{1}$. Considering the curves corresponding to charge
degeneracy i.e. $\Delta E_{C}=0$, we note that $n_{ph}$ is zero until
the bias voltage, $eV$ reaches $2\hbar\omega_{1}$, after which it
starts to increase continuously with bias voltage. For the curves
corresponding to $\Delta E_{C}=0.25\hbar\omega_{1}$ the onset occurs
at $eV=1.5\hbar\omega_{1}$ and there is an additional kink on each
curve at $eV=2.5\hbar\omega_{1}$.
\\
\\
These onsets and kinks can be understood from the energetics of
allowed tunneling processes: Due to the continuous density of states
of the dot all electrons in the lead with energies above $\Delta
E_{C}$, can tunnel into the dot.  Photon emission by the tunneling
electrons is however only possible for electrons with energies above
$\hbar\omega_{1}+\Delta E_{C}$. Similarly, an electron in the dot can
tunnel out to unoccupied states in the leads with energies below
$\Delta E_{C}$, but can only tunnel out with photon emission to states
with energies below $\Delta E_{C}-\hbar\omega_{1}$. Therefore, at low
temperatures photon emission is only possible by an electron tunneling
from (to) the left (right) lead for a bias voltages
$eV/2\geq\hbar\omega_{1}+(-)\Delta E_{C}/2$. The onsets and kinks in
Fig.~\ref{meanphoton1} thus correspond to thresholds of tunneling
processes with photon emission into the cavity.
\\
\\
The rate of increase of the population $n_{ph}$ with increasing $eV>
2\hbar\omega_{1}-\Delta E_{C}$ can most easily be understood for
$\Delta E_{c}=0$. We see in Fig.~\ref{meanphoton1} that the population
goes from growing almost linearly for $\lambda_{1}=0.2$ to a slower,
sub linear increase for larger $\lambda_{1}\sim 1$. In the limit
$\lambda_{1}\ll 1$ an analytical formula for the photon distribution,
$\{P_{n}\}$, can be derived by only taking into account processes to
leading order in $\lambda_{1}$ (see Appendix C). For
$eV-2\hbar\omega_{1}\gg k_{B}T$ we obtain
\begin{align}
&P_{n}=\frac{2\hbar\omega_{1}}{eV+2\hbar\omega_{1}}\left(\frac{eV-2\hbar\omega_{1}}{eV+2\hbar\omega_{1}}\right)^{n},
\label{distr}
\end{align}
independent of $\lambda_{1}$. We note that the probabilities, $P_{n}$, are
Boltzmann distributed. Hence the distribution can be described by an effective temperature
\begin{equation} 
k_{B}T_{\text{eff}}=\hbar\omega_{1}/\ln[(eV+2\hbar\omega_{1})/(eV-2\hbar\omega_{1})]. 
\label{Teff}
\end{equation}
Using standard thermodynamics we then obtain the following linear relation between 
the population and bias voltage as
\begin{align}
n_{ph}=\frac{1}{\exp\left(\hbar\omega_{1}/[k_{B}T_{\text{eff}}]\right)-1}=\frac{eV-2\hbar\omega_{1}}{4\hbar\omega_{1}}.  
\label{anamean}
\end{align} 
Looking at Fig.~\ref{meanphoton1} we see that $n_{ph}$ is well
described by Eq.~\eqref{anamean} for coupling strengths up to
$\lambda_{1}\approx 0.2$. The slower increase with voltage for larger
$\lambda_{1}$ can be understood as follows: In the limit
$\lambda_{1}\ll 1$ only processes where the number of photons are
changed $-1,0$ or $1$ are important, since they are the only ones
having non-zero amplitude to leading order in $\lambda_{1}$. This is
deduced from the corresponding Franck-Condon factors [see
Eq.~\eqref{Franck}]. However, at the considered bias voltages only
processes where the number of photons is \textit{increased} by at most
one are allowed energetically. Thus, when $\lambda_{1}$ is increased
the rate for the higher order processes where the photons number is
\textit{decreased} becomes larger, but not for the ones where the
photon number is increased. Hence, the population $n_{ph}$ is
decreased. The results are qualitatively similar for $\Delta
E_{C}=0.25\hbar\omega_{1}$.
\\
\\
To further investigate the properties of the distribution, $\{P_{n}\}$
for coupling strengths approaching $\lambda_{1}\sim 1$, $P_{n}$ is
plotted against $n$ for bias $eV=3\hbar\omega_{1}$ in Fig.~\ref{prob}.
We see that the distribution decreases exponentially with $n$ for
couplings $\lambda_{1}\ll 1$ in line with Eq.~\eqref{distr}. For
stronger couplings the decrease is faster, due to higher order
relaxation processes. This observation shows that the probabilities
$P_{n}$ are not Boltzmann distributed and hence an effective
temperature cannot be defined. The cavity mode is thus clearly in a
non-thermal state. This can be further illustrated by investigating
e.g. the photon Fano factor\cite{merlo08} (not presented here).
\\
\\
An important feature of the photon state, not captured in the above analysis,
is that an electron tunneling into the dot displaces the harmonic oscillator
corresponding to the first cavity mode by an amount proportional to the
coupling strength, $\lambda_{1}$. To illustrate the effect of the displacement
of the mode we plot in Fig.~\ref{Q} the Wigner-function\cite{cahill69}
\begin{equation}
W(\beta)=\int{\frac{d^{2}\xi}{\pi}}\tr\left(\rho_{s}\exp\left[\xi\hat{a}_{1}^{\dagger}-\xi^{*}\hat{a}_{1}\right]\right)\exp\left(\xi\beta^{*}-\xi^{*}\beta\right),
\label{wig}
\end{equation}
where the trace is taken over both electron and photon degrees of freedom.
From Fig. \ref{Q} we note that for coupling $\lambda_{1}=0.2$, we can only
discern a single peak of the Wigner function while for the larger coupling,
$\lambda_{1}=2$, the peak is split into two. The second peak comes from the photons of the polaron
of the charged dot and it becomes visible for coupling strengths
$\lambda_{1}\sim 1$. We also note that $\Delta E_{C}$ has an
impact on the photon distribution as the second peak is weaker for $\Delta
E_{C}=0.25\hbar\omega_{1}$ than for $\Delta E_{C}=0$. This is a consequence of
a smaller probability of the dot being occupied in the previous case. 
\\
\\
It is interesting to briefly compare our results to those obtained for
the current-induced non-equilibrium state of a single boson mode
coupled to a single-level, see
e.g. Refs. \onlinecite{Mitra04,hartle11,Koch06,hartle112}.  For a
single-level dot, in contrast to our metallic dot, the population
grows stepwise with bias voltage, where each step corresponds to an
onset of photon emission in a tunneling process. Furthermore, in
contrast our result Eq.~\eqref{anamean}, the photon distribution and
hence the population is not convergent for charge degeneracy, $\Delta
E_{C}=0$, in the limit of couplings, $\lambda_{1}\ll 1$, for voltages
above the first onset of photon emission. \cite{Koch06,hartle11} This
is because the rate for going from a state with $n$ to a state with
$n+1$ photons is equal to the rate for the opposite process, which
gives an equal probability of all photon states.  In metallic dot the
processes $n+1\rightarrow n$ has larger rate than $n\rightarrow n+1$,
as discussed in detail in Appendix C.
\\
\\
\subsection{Two active modes}

\begin{figure}[h]
\includegraphics[trim=0cm 0cm 0cm 0cm, clip=true, width=0.45\textwidth, angle=0]{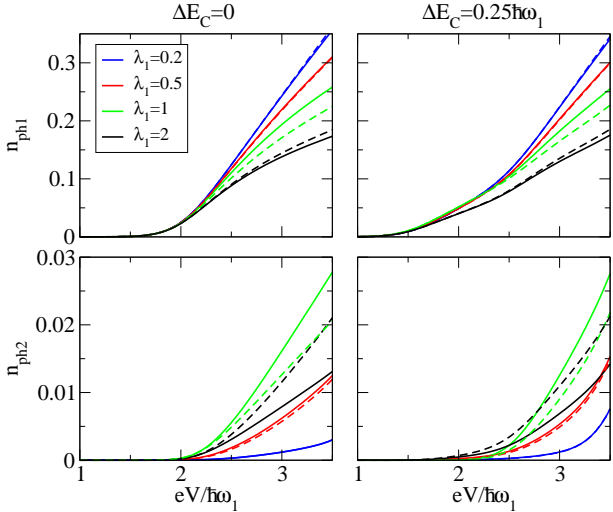}
  \caption{Color online). Mean number of photons in first and second
    mode, $n_{ph1}, n_{ph2}$, against bias voltage for different
    coupling strengths and $\Delta E_{C}$. The temperature is
    $k_{B}T=0.05\hbar\omega_{1}$. Solid (dashed) lines show results
    with (without) coherences retained.}
\label{meanphoton}
\end{figure}

We then turn to the case with two active modes with
$\lambda_{1}=\sqrt{2}\lambda_{2}$.  As for the single-mode case, we
first consider the average number of photon excitations in the two
polaron states, defined by $n_{ph1(2)}=\sum_{n,m}n(m)\left[\braket{ 0n
    m|\rho_{s}| 0 n m}+\braket{1n m|\rho_{s}| 1 n m}\right]$. The
dependence of $n_{ph1}$ and $n_{ph2}$ on bias voltage for different
coupling strengths are depicted in Fig.~\ref{meanphoton}. We see that
the onsets and slopes in the curves for $n_{ph1}$ show the same
qualitative behavior as in the single mode case. Moreover, importantly
$n_{ph2}$ have onsets and kinks at the same bias voltages. This is
despite the fact that direct excitation of this mode is not
energetically allowed at the considered bias voltages. The population
in the second mode is thus due to \textit{inter-mode conversion}. The
mechanism of this conversion is that a tunneling electron excites a
photon in the second mode and simultaneously de-excites a photon in the
first mode.  Since the change of the energy of the tunneling electron
is the same as when it emits a photon into the first mode, both
process become energetically allowed at the same bias voltage. We note
from Fig.~\ref{meanphoton} that $n_{ph2}$ initially increases with
$\lambda_{1}=1$ up to and starts to decrease again, for even large
$\lambda_{1}=2$. We also point out that there is a difference
between the results obtained from calculations with and without the
coherences retained.  This is particularly apparent for the coupling
strength $\lambda_{1}=1$. Here $n_{ph1}$ and $n_{ph2}$ are larger and
significantly larger, respectively, in the presence of coherences. To
identify the dependence of the coherent effects on the coupling
strengths we plot the difference between the coherent and the
incoherent occupations $n_{ph1}$ and $n_{ph2}$ as a function of
$\lambda_{1}$ and $\lambda_{2}$ for bias voltage $eV=3\hbar\omega_{1}$
in Fig.~\ref{diffcoher}.
\subsubsection*{Polaron coherences}

\begin{figure}[h]
\begin{center}
\includegraphics[trim=0cm 0cm 0cm 0cm, clip=true, width=0.45\textwidth, angle=0]{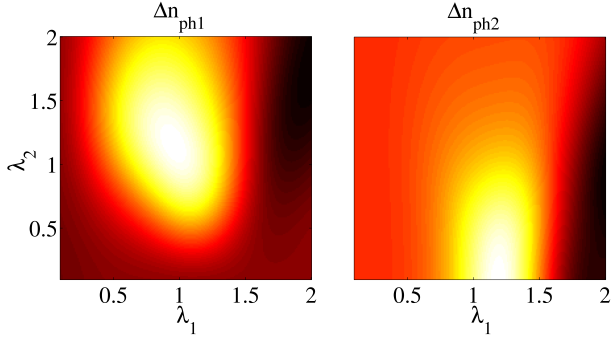}
\end{center}
\caption{(Color online) The difference between the obtained average
  number of excitations in the first mode $n_{ph1}$ (left) and second
  mode $n_{ph2}$ (right) when coherences are retained and not in the
  quantum master equation. The color scale goes from black (small
  difference) to white (large difference). The bias voltage is
  $eV=3\hbar\omega_{1}$ and temperature $k_{B}T=0.05\hbar\omega_{1}$.
}
\label{diffcoher}
\end{figure}

\begin{figure}[h]
\begin{center}
\resizebox{!}{45mm}{\includegraphics{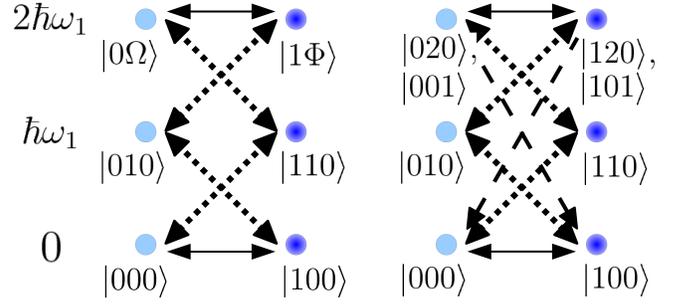}}
\end{center}
\caption{(Color online) Scheme of transition between different states
  for the effective master equation Eq. \eqref{anamaster} (left) and
  the corresponding incoherent master equation (right). The filled
  lines represents transitions when no photon is emitted or absorbed,
  the dotted lines represent transitions involving one photon. The
  dashed lines represents transitions where two photons are emitted.
}
\label{scheme}  
\end{figure}

As is clear from both Figs.~\ref{meanphoton},~\ref{diffcoher}, the
effect of coherences on $n_{ph1}$ and $n_{ph2}$ are most pronounced
around $\lambda_{1}=1$, for which they are enhanced. For the coupling
strength to the second mode, the effect of the coherence is maximal
around $\lambda_{2}\sim 1$ for $n_{ph1}$, while the effect on
$n_{ph2}$ is maximal for $\lambda_{2}\ll 1$. A representative pair of
couplings giving large coherence effects is $\lambda_{1}=1$ and
$\lambda_{2}=1/\sqrt{2}$.  For these specific coupling strengths a
detailed investigation of the coherences can be performed. By a
careful inspection of the numerically obtained steady-state density
matrix for voltages above onset we find that only a limited number of
polaron states have non-negligible amplitude. This allows us to
describe the long-time charge and photon properties by an effective
master-equation
\begin{equation}
\frac{d\bold{P}}{dt}=M\bold{P},
\label{anamaster}
\end{equation}
where $\bold{P}=[P_{0 0 0},\quad P_{0  1 0},\quad P_{0 \Omega},\quad P_{1 0
  0},\quad P_{1  1 0},\quad P_{1 \Phi}]^{T}$, with $P_{\mu 0 0}, P_{\mu 10}$, $P_{0
  \Omega}$ and $P_{1
  \Phi}$ being the probabilities for the states $\ket{\mu 0 0},\ket{\mu
1 0}$, $\ket{0 \Omega}$ and $\ket{1 \Phi}$, respectively. The latter states are given by
\begin{align}
\ket{0 \Omega}=\frac{\ket{0 2 0}-\ket{0 0 1}}{\sqrt{2}},\quad\ket{1 \Phi}=\frac{\ket{1 2 0}+\ket{1 0 1}}{\sqrt{2}}, 
\label{basis}
\end{align}
and are thus superpositions of degenerate polaron states with two photons in the
first mode and one photon in the second mode. The matrix $M$ in
Eq.~\eqref{anamaster} is further given by
\begin{align}
&M/\tilde{\Gamma}=\nonumber\\
&\!\!\!\!\!\!\!\!\!\!\!\!\!\begin{pmatrix}
-\underset{j=0,1}{\sum}\tilde{h}_{j} & 0 \!\!\! & 0\!\! &\tilde{g}_{0}  & \tilde{g}_{1} & \!\!\!\!\!\!\!\!\!\!\!0 \\
0 & -\underset{j=-1,1}{\sum}\!\!\!\!\tilde{h}_{j}\!\!\! &  0 &
\tilde{g}_{-1} & 0 &\!\!\!\!\!\!\!\!\!\!\!\tilde{g}_{1}\\
0 & 0\!\! & -\underset{j=-1,0}{\sum}\!\!\!\!\tilde{h}_{j}\!\!\! & 0\!\!&
\tilde{g}_{-1}  & \!\!\!\!\!\!\!\!\!\!\!\!\tilde{g}_{0}\\
\!\!\!\tilde{h}_{0} & \tilde{h}_{-1}\!\! & 0\!\! &
-\underset{j=0,-1}{\sum}\!\!\!\!\tilde{g}_{j}& 0 & \!\!\!\!\!\!\!\!\!\!0 \\
\tilde{h}_{1} & 0 & \tilde{h}_{-1} \!\! & 0 \!\!  &
-\underset{j=-1,1}{\sum}\!\!\!\!\tilde{g}_{j} & \!\!\!\!\!\!\!\!\!\!0 \\
0 & \tilde{h}_{1}  & \tilde{h}_{0}\!\! & 0 \!\!  & 0\!\! &\!\!\!\!\!\!\!\!-\underset{j=0,1}{\sum}\tilde{g}_{j}
\end{pmatrix},
\label{M}
\end{align}
where $\tilde{h}_{j}=\sum_{\ell=L,R}h_{\ell}(j\hbar\omega_{1}+\Delta E_{C})$
and $\tilde{g}_{j}=\sum_{\ell=L,R}g_{\ell}(j\hbar\omega_{1}+\Delta E_{C})$.
\\
\\
To illustrate the origin of the coherences in the master equation, the
transitions described by Eq.~\eqref{anamaster} are depicted in
Fig.~\ref {scheme}. For comparison, the transitions of the
corresponding incoherent master equation, with probabilities $P_{\mu
  00}$, $P_{\mu 10}$, $P_{\mu 2 0}$ and $P_{\mu 0 1}$, are shown to
the right. The transitions between states with an energy in the cavity
modes less or equal to $\hbar\omega_{1}$ are the same in the coherent
and incoherent case.  For transitions to, from or between states with
an energy $2\hbar\omega_{1}$ in the cavity modes, however, the picture
is different. Consider e.g. the transition from $\ket{1 1 0}$ in which
an additional quanta, $\hbar\omega_{1}$, is excited in the cavity
modes.  Two processes contribute to this transition: An additional
photon can be excited in the first mode, $\ket{1 1 0}\rightarrow
\ket{0 2 0}$, and a photon can be excited in the second mode by
inter-mode conversion $\ket{1 1 0}\rightarrow \ket{0 0 1}$.  Since
$\ket{0 2 0}$ and $\ket{ 0 0 1}$ are degenerate the final state is a
superposition of them in the coherent case, $\ket{0\Omega}$, while
there is no superposition in the incoherent case. Similar explanations
hold for the other transitions to, from or between states with energy
$2\hbar\omega_{1}$.
\\
\\
From Eq.~\eqref{anamaster} an expression for the steady-state density
matrix, $\rho_{s}$, can be obtained. Considering e.g. $\Delta E_{C}=0$ and $eV-2\hbar\omega_{1}\gg k_{B}T$ the steady-
state density matrix has the simple form 
\begin{align}
&\rho_{s}=\frac{1}{2}\frac{1}{3(eV)^{2}+4(\hbar\omega_{1})^{2}}\nonumber\\
&\times[\left(\ket{000}\bra{000}+\ket{100}\bra{100}\right)(eV+2\hbar\omega_{1})^{2}\nonumber\\
&+\left(\ket{010}\bra{010}+\ket{110}\bra{110}\right)\left[(eV)^2-4(\hbar\omega_{1})^{2}\right]\nonumber\\
&+[\ket{0\Omega}\bra{0 \Omega}+\ket{1 \Phi}\bra{1 \Phi}][eV-2\hbar\omega_{1}]^{2}].
\label{steady}
\end{align} 
This expression clearly shows that the superpositions of polaron
states have finite probabilities. We point out that $\ket{0 \Omega}$ is an
odd superposition while $\ket{1 \Phi}$ is even [see Eq.~\eqref{basis}]. 
The amplitudes for the two states in the superposition are determined by the
Franck-Condon factors for the two processes [see Eq.\eqref{gme}].
The steady-state density matrix thus displays non-trivial
correlations between the cavity photon state and the charge state of the dot. From
Eq. \eqref{steady} we also find the populations 
\begin{align}
&n_{ph1}=\frac{2eV(eV-2\hbar\omega_{1})}{3(eV)^{2}+4(\hbar\omega_{1})^{2}}\nonumber\\
&n_{ph2}=\frac{1}{2}\frac{(eV-2\hbar\omega_{1})^{2}}{3(eV)^{2}+4(\hbar\omega_{1})^{2}},
\end{align}
in good agreement with the numerical results in Fig.~\ref{meanphoton}.
\\
\\
To understand qualitatively the effect of the coherences on the
populations we again consider Fig.~\ref{scheme}. We see that the major
difference between the coherent and incoherent master equation is that
there is no direct relaxation from states with energy
$2\hbar\omega_{1}$ to states with zero energy in the photon modes in
the former case. This coherent blocking of relaxation provides a
plausible explanation of why $n_{ph1}$ and $n_{ph2}$ are enhanced for
this case (see Fig.~\ref{meanphoton}).

\section{Electronic transport and noise}
Having investigated the current-induced non-equilibrium photon state
we now turn to the properties of the electronic transport itself,
fully accounting for the back-action of the cavity photons on the
tunneling electrons. We focus our investigation on the average current
and the low frequency current fluctuations, or noise,
\cite{Blanterrev} experimentally accessible in metallic quantum
dots.\cite{Kafanov09} The current $I$ and the noise $S$ can
conveniently be calculated from the number resolved version of the
quantum master equation Eq.~\eqref{gme}, as discussed in the context
of full counting statistics, see e.g. early works
\cite{Bagrets03,Flindt05,Kiesslich06,Braggio05} for a detailed
discussion. For completeness of the present work we give in Appendix B
a short derivation of the expressions for the current and the noise,
used in the analytical and numerical calculations below.

\subsubsection*{Conductance and noise for a single active mode}
\begin{figure}[h]
\begin{center}
\includegraphics[trim=0cm 0cm 0cm -1cm, clip=true, width=0.45\textwidth, angle=0]{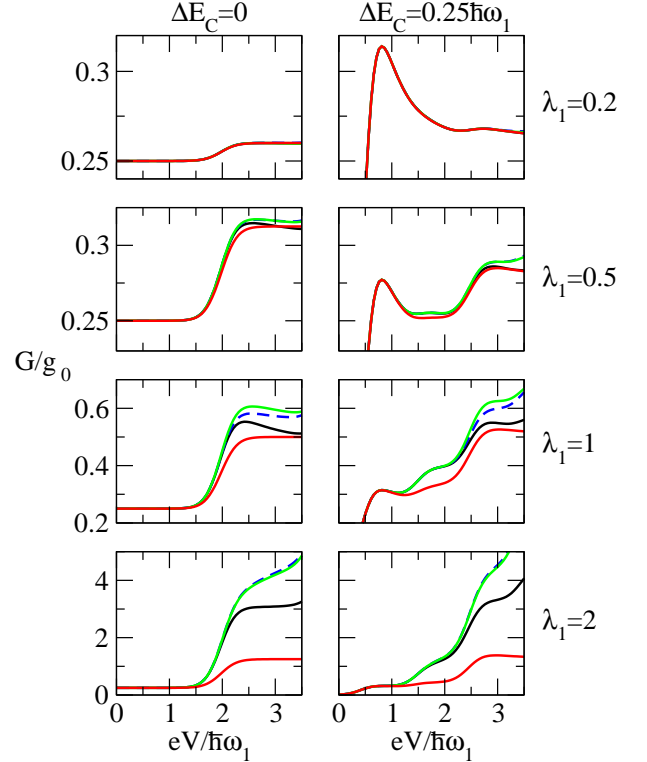}
\end{center}
\caption{(Color online) Differential conductance as a function of bias
  voltage, in units of
  $g_{0}=e^{2}\tilde{\Gamma}/\hbar\omega_{1}$. The results are
  obtained by a quantum master equation for one mode (black), two
  modes without coherences (dashed blue), two modes with coherences
  (green) and for an equilibrated photon distribution at temperature
  $k_{B}T_{ph}\ll \hbar\omega_{1}$ (red).  The electron temperature
  was $k_{B}T=0.05\hbar\omega_{1}$.  }
\label{diffcond}
\end{figure}

\begin{figure}[h]
\begin{center}
\includegraphics[trim=0cm 0cm 0cm -1cm, clip=true, width=0.45\textwidth, angle=0]{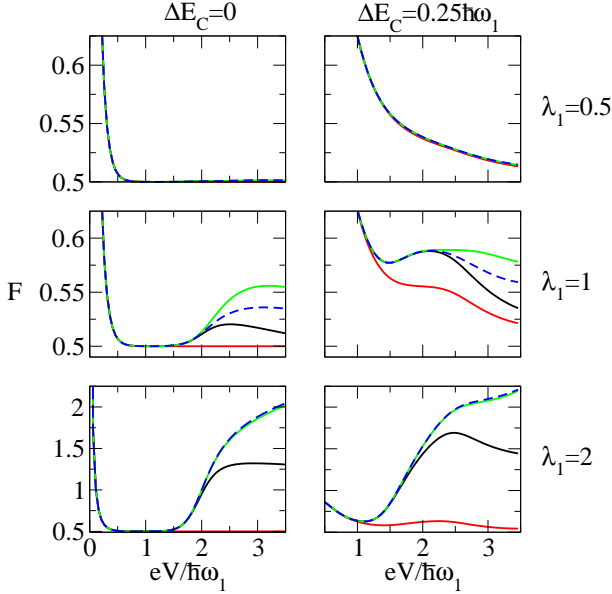}
\end{center}
\caption{(Color online) Fano-factor as a function of bias voltage obtained for one mode (black), two modes
  without coherences (dashed blue), two modes with coherences (green) and for the
  first mode being equilibrated (red). The temperature was $k_{B}T=0.05\hbar\omega_{1}$.} 
\label{noise}
\end{figure}

We first consider the I-V characteristics when only a single mode is
active, i.e. $\lambda_{2}=0$. In Fig.~\ref{diffcond} the conductance
$G=dI/dV$ is plotted against bias voltage for different coupling
strengths and charging energy differences $\Delta E_{C}$. The main
feature of the conductance is a stepwise increase as the bias voltage
passes $2\hbar\omega_{1}$ and $(2\pm 0.5)\hbar\omega_{1}$ for $\Delta
E_{C}=0$ and $\Delta E_{C}=0.25\hbar\omega_{1}$, respectively. As
concluded in the last section, at these bias voltages photon emission
in the tunneling process becomes energetically allowed. In the low
bias regime, $eV<2\hbar\omega_{1}-\Delta E_{C}$, the cavity modes
effect the transport only by renormalizing the tunneling rate
[Franck-Condon effect see Eq.~\eqref{renormeq}]. Considering
specifically $\Delta E_{C}=0$ the conductance is
\begin{equation}
G_{0}=\frac{e^{2}\tilde{\Gamma}}{4\hbar\omega_{1}}
\label{Gbelow}
\end{equation} 
for any $\lambda_{1}$. For $eV>2\hbar\omega_{1}$ the electrons can also tunnel by emitting or
absorbing a photon in the first mode. Thus, additional transport channels open up which gives the increase in
conductance.  For $\lambda_{1}\ll 1$ an analytical
formula can be derived for the conductance (see Appendix C). For bias voltages
$eV-2\hbar\omega_{1}\gg k_{B}T$, the conductance is given by
\begin{align}
G_{1}=\frac{e^{2}\tilde{\Gamma}(1+\lambda_{1}^{2})}{4\hbar\omega_{1}}.
\label{weakc}
\end{align}  
Thus, the contribution from the additional channels scales
as $\lambda_{1}^{2}$. This dependence derives from the rate
of emission or absorption of one photon in a tunneling event,
proportional to $|Y_{nn+1}^{1}|^{2}\propto \lambda_{1}^{2}$. Interestingly, the result in Eq.~\eqref{weakc} is
independent on the distribution $\{P_{n}\}$. For larger coupling
strengths $\lambda_{1}\sim 1$ processes of higher order in $\lambda_{1}$ start
to contribute to the conductance and Eq.~\eqref{weakc} no longer holds.
The rate of tunneling into and out of the dot will be dependent on the
number of photons in the cavity, i.e. the conductance becomes dependent on the
distribution $\{P_{n}\}$. As is seen in Fig.~\ref{diffcond}, the higher
order processes typically lead to an increased conductance.
\\
\\
To gain further insight into the effect of the coupling to the photon
mode on the electron transport properties, we investigate the
correlations between the tunneling electrons. The correlations are
quantified by the Fano-factor $F=S/(eI)$. For a dot decoupled from the
cavity the electrons are anti-correlated due to the Coulomb
interaction, and F is always less than one for bias voltage $eV-\Delta
E_{C}\gg k_{B}T$. The Fano-factor for the dot coupled to a single
photon mode is plotted against bias voltage in Fig.~\ref{noise}.
Below the onset voltage the only effect of the coupling between the
dot and the cavity mode is a renormalization of the tunneling rates.
Focusing on $\Delta E_{C}=0$, the noise is
\begin{equation}
S=\frac{e^{3}V\tilde{\Gamma}}{8\hbar\omega_{1}},
\label{Bonsetnoise}
\end{equation}
giving a Fano-factor $1/2$. Above onset, i.e. for bias voltages
$eV-2\hbar\omega_{1}\gg k_{B}T$, the noise becomes dependent on the
coupling strength $\lambda_{1}$. In the limit $\lambda_{1}\ll 1$ an
expression for the noise can be found analytically (See Appendix
C). We find
\begin{equation}
S=\frac{e^{2}\tilde{\Gamma}}{4\hbar\omega_{1}}\left(\frac{eV(1+\lambda_{1}^{2})}{2\hbar\omega_{1}}-\lambda_{1}^{2}\right)=\frac{eI}{2}
\label{lowlimitnoise}
\end{equation} 
which gives a Fano-factor of $1/2$ above the onset voltage as
well. Thus, the onset of photon emission does not change the
correlations between the tunneling electrons. We point out that
corrections to the Fano-factor in Eq.~\eqref{lowlimitnoise} is of
order $\lambda_{1}^{4}$. Consequently, as can be seen in
Fig.~\ref{noise}, the deviation in the Fano-factor from $1/2$ for
$\Delta E_{C}=0$ is small even for coupling strengths as large as
$\lambda_{1}=0.5$. However, for coupling strengths approaching
$\lambda_{1}\sim 1$ we see an increase in the Fano-factor as the bias
voltage passes $2\hbar\omega_{1}$ and for $\lambda_{1}=2$ we even get
super-poissonian noise. Similarly we see for $\Delta
E_{C}=0.25\hbar\omega_{1}$ that there is an increase in the
Fano-factor for the bias voltage $1.5\hbar\omega_{1}$ for coupling
strengths $\lambda_{1}\sim 1$.
\\
\\
Thus, the change in the Fano-factor above the onset voltage occurs for
coupling strengths deep into the ultrastrong coupling regime,
$\lambda_{1}\sim 1$.  To understand this we recall that the tunneling
into and out of the dot is dependent on the photon state for these
coupling strengths. For e.g. two subsequently tunneling electrons this
means that the tunneling rate for the later electron depends on which
photon state the cavity mode was left in by the first electron. For
the parameter regime investigated, this leads to an increased tendency
of bunching, and hence a larger Fano-factor. In an equivalent physical
picture the increase in the Fano-factor can be attributed to the
emergence of the avalanche effect found for a single level strongly
coupled to a boson mode described in Ref. \onlinecite{Koch05}. We thus
find that the effect is also present for a metallic dot coupled to a
boson mode.
\\
\\
To highlight the effect of the non-equilibrium photon distribution on
the transport properties it is instructive to compare the result
presented above to ones where the cavity modes are equilibrated at
$T_{ph}$. (see Appendix D for details). To keep the discussion short
we focus the discussion on temperatures $T_{ph}$ for which only the
first mode can have a finite thermal population and $\Delta
E_{C}=0$. We consider first the conductance $G^{th}$ and restate that
for coupling strengths $\lambda_{1}\ll 1$ the conductance is
independent on the photon distribution. Hence the conductance for an
equilibrated mode is given by Eqs.~\eqref{Gbelow} and~\eqref{weakc}
below and above the onset voltage, respectively.  For larger coupling
strengths, $\lambda_{1}\sim 1$, when the photon distribution affects
the transport, the conductances for equilibrated and non-equilibrated
modes differ. For bias voltages below the onset voltage the
conductance is given by
\begin{align}
&G^{th}_{0}=\frac{e^{2}\tilde{\Gamma}}{4\hbar\omega_{1}}\exp\left[-\lambda_{1}^{2}\left(\coth\left(\frac{\hbar\omega_{1}}{2k_{B}T_{ph}}\right)-1\right)\right]\nonumber\\
&\times \sum_{n=0}^{\infty}(2-\delta_{n0})\exp\left[\frac{-n\hbar \omega_1}{2k_BT_{ph}}\right]I_n\left[\lambda_{1}^{2}\sinh^{-1}\left(\frac{\hbar\omega_{1}}{2k_{B}T_{ph}}\right)\right].
\label{PEbelow}
\end{align}
Here $I_{n}$ denotes the n:th order modified Bessel function of the first
kind. The conductance in Eq.~\eqref{PEbelow} is an increasing function of the
temperature and it is thus larger than than the
conductance in Eq.~\eqref{Gbelow}. For bias voltages above the onset voltage
the conductance for the equilibrated mode is given by
$G^{th}_{1}=G^{th}_{0}+\Delta G^{th}$ with 
\begin{align}
&\Delta G^{th}=\frac{e^{2}\tilde{\Gamma}}{4\hbar\omega_{1}}\exp\left[-\lambda_{1}^{2}\left(\coth\left(\frac{\hbar\omega_{1}}{2k_{B}T_{ph}}\right)-1\right)\right] \nonumber \\
&\times 2\sinh\left(\frac{\hbar\omega_{1}}{2k_{B}T_{ph}}\right)I_1\left[\lambda_{1}^{2}\sinh^{-1}\left(\frac{\hbar\omega_{1}}{2k_{B}T_{ph}}\right)\right]
\label{PEabove}
\end{align}
For $k_{B}T_{ph}\ll\hbar\omega_{1}$ the expression for $G^{th}_{1}$
reduces to Eq.~\eqref{weakc}, obtained for a non-equilibrium photon
distribution in the limit $\lambda_{1}\ll 1$.
\\
\\
It can be shown (see appendix D) that $\Delta G^{th}/G_{0}^{th}$ is
limited by the low-temperature value $\lambda_{1}^{2}$.  Importantly,
for the non-equilibrium photon mode investigated above the relative
difference in conductance $\Delta G/G_{0}$ in Fig.~\ref{diffcond} is
not limited to $\lambda_{1}^{2}$.  To clearly illustrate the
difference between the conductances for a thermalized and a
non-equilibrium mode, $G^{th}$, is plotted as a reference in
Fig.~\ref{diffcond}.
\\
\\
Further insight is obtained by comparing the Fano-factors for an
equilibrated and non-equilibrated mode. For an equilibrated mode we
find (see Appendix D) that in the low-temperature limit,
$k_{B}T_{ph}\ll \hbar\omega_{1}$, for arbitrary couplings
$\lambda_{1}$ and $\Delta E_{C}=0$, the noise is given by
Eqs.~\eqref{Bonsetnoise} and~\eqref{lowlimitnoise} below and above the
onset voltage, respectively. The low-temperature Fano-factors are
plotted as a reference in Fig.~\ref{noise}.  For finite temperatures
the expressions for the noise below and above the onset voltage are
lengthy and do not provide additional physical insight. We therefore
simply provide the qualitative result: The Fano-factor decays
monotonically with bias voltage for a given $T_{ph}$. As is clear from
Fig.~\ref{noise} and the discussion above the later result is in
contrast to what we find for a non-equilibrium photon
distribution. The increase in the Fano-factor at the onset voltage for
ultrastrong couplings $\lambda_{1}\sim 1$ is thus a clear signature of
a non-equilibrium photon distribution of the cavity mode.

\subsubsection*{Conductance and noise for two active modes}
We then turn to the transport properties for the case with two active
modes, with couplings $\lambda_{1}=\sqrt{2}\lambda_{2}$.  The
differential conductance and the Fano-factor are plotted against bias
in Figs.~\ref{diffcond} and~\ref{noise}, for both the cases with and
the cases without coherences retained in the quantum master equation.
As for a single active mode there is a stepwise increase in
differential conductance as the bias voltage approaches
$2\hbar\omega_{1}$ and $2\hbar\omega_{1}\pm 0.5\hbar\omega_{1}$ for
$\Delta E_{C}=0$ and $\Delta E_{C}=0.25\hbar\omega_{1}$,
respectively. We note that the conductance is typically larger than
for the single mode case for a given coupling strength
$\lambda_{1}$. Thus the inter-mode conversion, discussed in the last
section, typically increases the conductance. Similarly, there is an
increase in the Fano-factor at the onset voltage.
\\
\\
We note that there is a difference between the conductance obtained
when coherences are included in the master equation and not. The
difference is most apparent for $\lambda_{1}=1$, where they lead to
enhancement of the conductance. This agrees with the finding that
$n_{ph1}$ and $ n_{ph2}$ show the most pronounced effect of the
coherences around this coupling strength (depicted in
Figs.~\ref{meanphoton} and~\ref{diffcoher}). We recall from the
previous section that processes where the energy in the photon modes
is decreased by more than $\hbar\omega_{1}$ are blocked when
coherences are retained in the master equation for $\lambda_{1}=1$. We
attribute the conductance enhancement to this blocking effect since
the blocked processes contribute to transfer of electrons in the
opposite direction to the applied bias.  We also see that the effect
of the coherences on the Fano-factor shows the most pronounced effect
at coupling strengths $\lambda_{1}\sim 1$. It is interesting to note
that in a very recent work on nanoelectromechanical systems,\cite{Yar}
the conductance of a few-level quantum dot coupled to several
vibrational modes was investigated, incorporating the effects of
coherence between degenerate vibrational states. 

\section{Conclusions}
In conclusion, we have investigated theoretically the properties of a
metallic quantum dot strongly coupled to a superconducting
transmission line cavity. The focus of the investigations has been on
the interplay between the cavity photon state and the electronic
transport through the dot. Based on the Lagrangian formulation of
circuit QED, a Hamiltonian for the system was
derived for arbitrary strong dot-cavity coupling. The electronic
transport and the photon dynamics were described by a quantum master
equation, fully accounting for coherent and non-equilibrium photon
effects. The cases with one and two active photon modes were
investigated. For a single active mode strongly coupled to the
conduction electrons, the photon state was found to be non-equilibrium,
with clear signatures of microwave polaron formation. For two active
modes coherence and photon conversion between the two modes was found.
Turning to the transport, the effect of the non-equilibrium photon
state on the electronic conduction was investigated by comparing to
the results for an equilibrated photon mode. Clear transport
signatures due to the non-equilibrium photon distribution were found,
in particular super-poissonian shot noise for strong dot-cavity
couplings.

\section*{Acknowledgments}
We thank G\"{o}ran Johansson, Takis Kontos, Per Delsing, Klaus
Ensslin, Kohnrad Lehnert and Olov Karlstr\"{o}m for fruitful
discussions and input. The work was supported by the Swedish VR. We
also thank Federica Haupt, Maura Sassetti, Fabio Cavaliere, Gianluca
Rastelli and Christian Flindt for constructive comments on an earlier
version of the manuscript.

\section*{Appendix A}

The time-evolution of the system is given by the Liouville equation
$\partial_{t}\hat{\rho}=-\frac{i}{\hbar}[H_{T}(t),\hat{\rho}(t)]$,
where $\hat{\rho}$ is the interaction picture density operator of the
system.  For weak tunnel coupling considered here we can restrict the
analysis to the sequential tunneling regime (Born approximation). We
first expand the Liouville equation to second order in the tunnel
coupling giving
\begin{equation}
\frac{d\hat{\rho}}{dt}=-\frac{i}{\hbar}[H_{T}(t),\hat{\rho}(\tau)]-\frac{1}{\hbar^{2}}\int_{\tau}^{t}dt'[H_{T}(t),[H_{T}(t'),\hat{\rho}(t)]]. 
\label{Liouville}
\end{equation}
Then the decoupled density operator 
$\hat{\rho}=\hat{\rho}_{L}\otimes\hat{\rho}_{O}\otimes\hat{\rho}_{S}$ is inserted.
Here $\hat{\rho}_{L}$, $\hat{\rho}_{O}$ and $\hat{\rho}_{S}$ are the
density operators of the leads, the fermionic degrees of freedom of the dot
and the dot charge-cavity system, respectively. Taking the dot and the leads to be in thermal equilibrium we can trace
Eq.~\eqref{Liouville} over the lead and fermionic dot degrees of freedom. 
Further, performing a Markov approximation and letting $\tau\rightarrow -\infty$ equation Eq.~\eqref{gme} is obtained for the matrix elements of
the Schr\"{o}dinger picture reduced density operator, $\rho$, in the polaron basis

\section*{Appendix B}

The starting point for the derivation of the current and and low
frequency noise is the expression for the cumulant generating function
$F(\chi)$. The cumulant generating function is given by the logarithm
of the Fourier transform of the distribution of probabilities $P(N,t)$
to transfer $N$ electrons through the dot during a measurement time
$t$, as $F(\chi)=-\ln(\sum_NP(N,t)\exp[iN\chi])$. The different
cumulants of the charge transfer are obtained by successive
differentiation of $F(\chi)$ with respect to the counting field
$\chi$. The first two cumulants are the current $I$ and noise $S$,
given by $I=(e/t)(-i\partial_{\chi}) F(\chi)|_{\chi=0}$ and
$S=(e^2/t)(-i\partial_{\chi})^2 F(\chi)|_{\chi=0}$, respectively.
\\
\\
To arrive at $F(\chi)$ in our model we first write the
$N$-resolved version of the quantum master equation in Eq.~\eqref{gme}
on a vectorized form. Fourier transforming with respect to $N$ we then
get the equation
$d\boldsymbol{\rho}(\chi)/dt=M(\chi)\boldsymbol{\rho}(\chi)$. The
cumulant generating function is given by the eigenvalue of $M(\chi)$
that goes to zero for $\chi=0$. For our purposes, to obtain explicit
expressions for the different cumulants, the generating function can
conveniently be written as the solution to the eigenvalue equation
\begin{equation}
M(\chi)\boldsymbol{\rho}(\chi)=F(\chi)\boldsymbol{\rho}(\chi).
\label{CGF}
\end{equation}
We then expand all quantities in $\chi$ as
$F(\chi)=(i\chi/e)I+(i\chi/e)^2S/2+..$, $M(\chi)=M_0+i\chi M_1+...$
and $\boldsymbol{\rho}(\chi)=\boldsymbol{\rho}^{(0)}+i\chi
\boldsymbol{\rho}^{(1)}+..$, which inserted into Eq.~\eqref{CGF} gives a hierarchy of
coupled linear equations as
\begin{eqnarray}
&&M^{(0)}\boldsymbol{\rho}^{(0)}=0, \hspace{0.5cm}
M^{(0)}\boldsymbol{\rho}^{(1)}+M^{(1)}\boldsymbol{\rho}^{(0)}=I\boldsymbol{\rho}^{(0)}, \\ \nonumber
&&M^{(0)}\boldsymbol{\rho}^{(2)}+M^{(1)}\boldsymbol{\rho}^{(1)}+M^{(2)}\boldsymbol{\rho}^{(0)}=I\boldsymbol{\rho}^{(1)}+S\boldsymbol{\rho}^{(0)}/2,
...
\label{xiexpansion}
\end{eqnarray}
The zeroth order equation gives the steady state density matrix,
$\boldsymbol{\rho}^{(0)}$. Expressions for the higher order
$\boldsymbol{\rho}^{(n)}$ are obtained by combining the n:th and lower
order equations.  By multiplying the first and higher order equations
from the left with the left zero eigenvector ${\bf v}$ of $M^{(0)}$,
defined from ${\bf v}^TM^{(0)}=0$, inserting the expression for
$\bold{\rho}^{(n)}$ and imposing the normalization condition ${\bf
  v}^T\boldsymbol{\rho}^{(0)}=1$, the different cumulants are
obtained. These equations are then solved numerically and in some
limiting cases analytically (see e.g. Appendix C). For the numerical
evaluation it is convenient to follow Ref. \onlinecite{Flindt04} and
fix the single free parameter in $\boldsymbol{\rho}^{(n)}$, the
component parallel to $\boldsymbol{\rho}^{(0)}$, by imposing a
suitable normalization of $\boldsymbol{\rho}^{(n)}$. Formally, the
first two cumulants, current and noise, can be written as
\cite{Flindt04}
\begin{eqnarray}
I&=&e{\bf v}^TM^{(1)}\boldsymbol{\rho}^{(0)}\nonumber\\
S&=&eI-2e^2{\bf v}^TM^{(1)}RM^{(1)}\boldsymbol{\rho}^{(0)}
\label{currnoise}
\end{eqnarray}
where $R$ denotes the pseudo-inverse of the singular matrix $M^{(0)}$ and we used that
$M^{(2)}=M^{(1)}/2$.
\section*{Appendix C}
We here present the derivation of analytical formulas for the photon
distribution, the current and the noise for a single cavity mode
coupled to the dot in the limit $\lambda_{1}\ll 1$, for charge
degeneracy, $\Delta E_{c}=0$. Performing the secular approximation on
Eq.~\eqref{gme} the following standard master equation, including
counting fields (see appendix B), is obtained
\begin{align}
\sum_{\ell}\left(\begin{array}{cc}
M^{00}_{\ell} & M^{10}_{\ell}e^{i\chi_{\ell}}\\
M^{01}_{\ell}e^{-i\chi_{\ell}} &M^{11}_{\ell}
\end{array}\right)\left(\begin{array}{c}
\bold{P}_{0}(\chi)\\
\bold{P}_{1}(\chi)\end{array}\right)=
F(\chi)\left(\begin{array}{c}
\bold{P}_{0}(\chi)\\
\bold{P}_{1}(\chi)\end{array}\right),
\label{singlemodeCMF}
\end{align}
where $\ell =L,R$, $\chi_{L}=0$ and $\chi_{R}=\chi$.  Here
$\bold{P}_{0}(\chi)=\left[\braket{0 0 0|\rho(\chi)|0 0 0}, \braket{0 1
    0|\rho(\chi)|0 1 0} .....\right]^{T}$ and
$\bold{P}_{1}(\chi)=\left[\braket{1 0 0|\rho(\chi)|1 0 0}, \braket{1 1
    0|\rho(\chi)|1 1 0} .....\right]^{T}$ are vectors corresponding to
$0$ or $1$ electrons on the dot, and the elements of the
$M_{\ell}-$matrices are given by
\begin{align}
&(M^{00}_{\ell})_{nm}=-\delta_{nm}\tilde{\Gamma}\sum_{k=-1}^{1}|Y_{n(n+k)}^{1}|^{2}h_{\ell}(k\hbar\omega_{1})\nonumber\\
&(M^{11}_{\ell})_{nm}=-\delta_{nm}\tilde{\Gamma}\sum_{k=-1}^{1}|Y_{n(n+k)}^{1}|^{2}g_{\ell}(k\hbar\omega_{1})\nonumber\\
&(M^{10}_{\ell})_{nm}=\tilde{\Gamma}\sum_{k=-1}^{1}\delta_{n(m-k)}|Y_{n(n+k)}^{1}|^{2}g_{\ell}(k\hbar\omega_{1})\nonumber\\
&(M^{01}_{\ell})_{nm}=\tilde{\Gamma}\sum_{k=-1}^{1}\delta_{n(m-k)}|Y_{n(n+k)}^{1}|^{2}h_{\ell}(k\hbar\omega_{1}),
\end{align}
where the renormalized Franck-Condon factors $Y^{1}_{nm}$ are defined
below Eq.~\eqref{renorm}. Here terms up to second order in
$\lambda_{1}$ are kept in $|Y_{nm}^{1}|^{2}$ (only $|Y_{nm}^{1}|^{2}$ with
$m=n,n\pm 1$ contribute). By expanding Eq.~\eqref{singlemodeCMF} to
zeroth order in $\chi$ the equation for the steady state probabilities
$\bold{P}^{(0)}_{0}$ and $\bold{P}^{(0)}_{1}$ are recovered. Since
$h_{R}(x)=g_{L}(-x)$ and $h_{L}(x)=g_{R}(-x)$ we have
$M^{00}_{L(R)}=M^{11}_{R(L)}$ and $M^{01}_{L(R)}=M^{10}_{R(L)}$. The
equations for $\bold{P}^{(0)}_{0}$ and $\bold{P}^{(0)}_{1}$ are thus
symmetric and we can write
$\bold{P}^{(0)}_{0}=\bold{P}_{1}^{(0)}=\bold{P}^{(0)}$ and obtain the
following equation for $P_{n}^{(0)}=(\bold{P}^{(0)})_n$ as
\begin{align}
\sum_{\ell}\left[-\left[|Y_{nn-1}^{1}|^{2}h_{\ell}(-\hbar\omega_{1})+|Y_{nn+1}^{1}|^{2}h_{\ell}(\hbar\omega_{1})\right]P^{(0)}_{n} \right.\nonumber\\
\left.+|Y_{nn+1}^{1}|^{2}h_{\ell}(-\hbar\omega_{1})P^{(0)}_{n+1}+|Y_{nn-1}^{1}|^{2}h_{\ell}(\hbar\omega_{1})P^{(0)}_{n-1}\right]=0,
\label{P0eq}
\end{align}
which has the solution 
\begin{equation}
P^{(0)}_{n}=\frac{(1-\eta)\eta^{n}}{2}
\end{equation}
with
$\eta=\left[\sum_{\ell}h_{\ell}(\hbar\omega_{1})/\sum_{\ell}h_{\ell}(-\hbar\omega_{1})\right]$
and where we have imposed the normalization condition
$2\sum_{n}P^{(0)}_{n}=1$. We point out that despite the expression
being independent on $\lambda_{1}$ it is correct to order
$\lambda_{1}^{2}$. For $eV-2\hbar\omega_{1}\gg k_{B}T$ we have
$\eta=(eV-2\hbar\omega_{1})/(eV+2\hbar\omega_{1})$ giving
Eq.~\eqref{distr}.  (Note that in the main text we use $P_{n}$ for
$P^{(0)}_{n}$ for notational convenience).  We also note that $\eta$
is the ratio between the rates of electron tunneling with photon
emission and tunneling with photon absorption. This ratio is always
smaller than one, which ensures that the distribution is convergent.
\\
\\
The current is calculated according to Eq. \eqref{currnoise}. For
$eV-2\hbar\omega_{1}\gg k_{B}T$ this gives
\begin{align}
&I=e\tilde{\Gamma}\sum_{n}\left[\underbrace{n\lambda_{1}^{2}\frac{eV+2\hbar\omega_{1}}{2\hbar\omega_{1}}}_{\Gamma_{n\uparrow}/\tilde{\Gamma}}+\underbrace{(1-2n\lambda_{1}^{2})\frac{eV}{2\hbar\omega}}_{\Gamma_{n0}/\tilde{\Gamma}}\right.\nonumber\\
&\left.+\underbrace{(n+1)\lambda_{1}^{2}\frac{eV-2\hbar\omega_{1}}{2\hbar\omega_{1}}}_{\Gamma_{n\downarrow}/\tilde{\Gamma}}\right]P^{(0)}_{n}=\frac{e\tilde{\Gamma}}{2}\left(\frac{eV(1+\lambda_{1}^{2})}{2\hbar\omega_{1}}-\lambda_{1}^{2}\right).
\label{curr}
\end{align}
From this equation Eq.~\eqref{weakc} follows directly. Furthermore,
the expression allows us to identify the contributions
$\Gamma_{n\downarrow}$, $\Gamma_{n0}$ and $\Gamma_{n\uparrow}$ to the
total rate for tunneling into/out of the dot in a state with $n$
photons.  We see from Eq.~\eqref{curr} that the rates for absorbing,
$\Gamma_{n\downarrow}$, or emitting a photon, $\Gamma_{n\uparrow}$ in
the tunneling process increases with $n$. This increase is however
canceled by an equally large decrease in the rate for tunneling
without photon emission or absorption, $\Gamma_{n0}$. This
cancellation makes the effective rate independent of $n$. The
current will therefore be independent on the distribution
$\{P^{(0)}_{n}\}$.
\\
\\
The noise can most conveniently be obtained from the expression for
the generating function $F(\chi)$. Above onset, for
$eV-2\hbar\omega_1\gg k_BT$, there is no tunneling against the bias
and the matrices $M_L^{10}, M_R^{01}, M_L^{11}$ and $M_R^{00}$ in
Eq. (\ref{singlemodeCMF}) can be neglected. Using the symmetries of
the $M_{\ell}$ matrices we can then write Eq. (\ref{singlemodeCMF}) as
\begin{align}
\left(\begin{array}{cc}
M^{00}_{R} & M^{10}_{R}e^{i\chi}\\
M^{10}_{R} &M^{00}_{R}
\end{array}\right)\left(\begin{array}{c}
\bold{P}_{0}(\chi)\\
\bold{P}_{1}(\chi)\end{array}\right)=
F(\chi)\left(\begin{array}{c}
\bold{P}_{0}(\chi)\\
\bold{P}_{1}(\chi)\end{array}\right).
\label{FCSred}
\end{align}
From Eq. (\ref{curr}) together with the expression for the current in
Eq. (\ref{currnoise}) it is clear that $e\tilde
v^TM_R^{10}\bold{P}^{(0)}=2I\tilde v^T\bold{P}^{(0)}=I$ where $\tilde
v^T=[1,1,1,...]$ and the normalization condition $\tilde
v^T\bold{P}^{(0)}=1/2$. Since the current is independent on $\bold{P}^{(0)}$
we have $\tilde v^TM_R^{10}=(2I/e)\tilde v^T$, i.e. $\tilde v^T$ is
the left eigenvector to $M_R^{10}$ with eigenvalue $2I/e$. Moreover,
from Eq. (\ref{P0eq}) for $\bold{P}^{(0)}$ we can write $\tilde
v^T(M^{00}_{R}+M_R^{10})=0$, i.e. $\tilde v^TM_R^{00}=-\tilde
v^TM_R^{10}=-(2I/e)\tilde v^T$. Multiplying both sides of
Eq. (\ref{FCSred}) from the left with $[\tilde v^T,\tilde v^T]$ then
gives
\begin{align}
\frac{2I}{e}\left(\begin{array}{cc}
-1 & e^{i\chi}\\
1 & -1
\end{array}\right)\left(\begin{array}{c}
\tilde v^T\bold{P}_{0}(\chi)\\
\tilde v^T \bold{P}_{1}(\chi)\end{array}\right)=
F(\chi)\left(\begin{array}{c}
\tilde v^T\bold{P}_{0}(\chi)\\
\tilde v^T\bold{P}_{1}(\chi)\end{array}\right).
\label{FCStwobytwo}
\end{align}
This $2\times2$ eigenvalue equation is directly solved, giving the
cumulant generating function
\begin{equation}
F(\chi)=\frac{2I}{e}\left(e^{i\chi/2}-1\right).
\label{CGFweak}
\end{equation}
From this expression we have, following Appendix B, the current $I$
and the noise $S=eI/2$, the expression in Eq. (\ref{lowlimitnoise}).

\section*{Appendix D}
 
We here present how the conductance and noise are calculated in the
case of equilibrated cavity modes at a temperature $T_{ph}$. Most of
the results presented in this section are available in the existing
literature.\cite{Nazarov92} They are included here merely for
completeness of the paper and to facilitate the comparison to the
non-equilibrium case.

The starting point for obtaining the conductance and noise for
thermally equilibrated modes is to derive a master equation for the
charge degree of freedom only. This derivation is to a large part
identical to the one presented in Appendix A. However, the density
operator $\hat{\rho}_{S}$ in Eq. \eqref{Liouville} is assumed to
factorize into $\hat \rho_{D}\otimes\hat \rho_{ph}$, where $\hat
\rho_{D}$ and $\hat \rho_{ph}$ are the density operators of the charge
degree of freedom and the thermally distributed photons,
respectively. Further, additional partial trace is taken over the
photon degrees of freedom. The following master equation for the
diagonal elements $P_0$ and $P_1$ of $\hat \rho_D$ is then obtained:
\begin{align}
\frac{d}{dt}\left(\begin{array}{c}
P_{0}\\
P_{1}
\end{array}\right)=\left(\begin{array}{cc}
-\Gamma_{01} & \Gamma_{10}\\
\Gamma_{01}  & -\Gamma_{10}\end{array}\right)\left(\begin{array}{c}
P_{0}\\
P_{1}
\label{MEPE}
\end{array}\right).
\end{align}
The rates $\Gamma_{01}=\Gamma_{01}^{+}+\Gamma_{01}^{-}$ and
$\Gamma_{10}=\Gamma_{10}^{+}+\Gamma_{10}^{-}$, where $\Gamma_{01(10)}^{\pm}$ is
the rate to tunnel in (+) or opposite to (-) the direction of the applied bias,
from 0 to 1 (1 to 0) excess charges on the dot, given by 
\begin{align}
\Gamma_{01(10)}^{\pm}=\frac{\tilde{\Gamma}}{\hbar\omega_{1}}\int_{-\infty}^{\infty}\!\!\!\!dEdE'f(E)[1-f(E')]\nonumber\\
\times\tilde{P}(E-E'\pm\frac{eV}{2}-(+)\Delta E_{C})
\label{PErates}
\end{align}
where  $\tilde{P}(E)=\exp(\sum_{p}\lambda_{p}^{2})P(E)$ and
\begin{equation}
P(E)=\frac{1}{2\pi\hbar}\int_{-\infty}^{\infty}dt\exp\left(\frac{iEt}{\hbar}\right)\prod_{p}\braket{[\hat{X}^{p}(t)]^{\dag}\hat{X}^{p}(0)}, 
\end{equation}
with 
\begin{equation}
\hat{X}^{p}(t)=\exp\left[-\lambda_{p}(\hat{a}_p^{\dag}e^{i\omega t}-\hat{a}_{p}e^{-i\omega t})\right].
\end{equation}
The function $P(E)$ is interpreted as the probability for an electron
to emit a net energy $E$ in to the cavity modes in the tunneling
event. This approach for studying tunneling in the presence of an
equilibrated electromagnetic environment is commonly referred to as
$P(E)$-theory.\cite{Nazarov92} $\tilde{P}(E)$ can be written
\begin{equation}
\tilde{P}(E)=\sum_{\{n_{p}\}}\delta(E-\sum_{p}n_{p}\hbar\omega_{p})\prod_{p}\tilde{P}^{p}_{n_{p}},
\end{equation}
with
\begin{align}
\tilde{P}^{p}_{n}=&\exp\left[\frac{n\hbar\omega_{p}}{2k_{B}T_{ph}}-\lambda_{p}^{2}\left(\coth\left(\frac{\hbar\omega_{p}}{2k_{B}T_{ph}}\right)-1\right)\right]\nonumber\\
\times& I_{n}\left[\lambda_{p}^{2}/\sinh(\hbar\omega_{p}/[2k_{B}T_{ph}])\right],
\end{align}
where $I_{n}$ is the n:th order modified Bessel function
of the first kind.\cite{Nazarov92}
\\
\\
From Eq.~\eqref{MEPE} the current and noise can now be obtained from
Eq.~\eqref{CGF} as
\begin{align}
I^{th}=&\frac{e(\Gamma_{10}^{+}\Gamma_{01}^{+}-\Gamma_{10}^{-}\Gamma_{01}^{-})}{\Gamma_{01}^{+}+\Gamma_{01}^{-}+\Gamma_{10}^{-}+\Gamma_{10}^{+}},\nonumber\\
S^{th}=&\frac{e^{2}(\Gamma_{10}^{+}\Gamma_{01}^{+}+\Gamma_{10}^{-}\Gamma_{01}^{-})}{\Gamma_{01}^{+}+\Gamma_{01}^{-}+\Gamma_{10}^{-}+\Gamma_{10}^{+}}-\frac{2e^{2}(\Gamma_{10}^{+}\Gamma_{01}^{+}-\Gamma_{10}^{-}\Gamma_{01}^{-})^{2}}{(\Gamma_{01}^{+}+\Gamma_{01}^{-}+\Gamma_{10}^{-}+\Gamma_{10}^{+})^{3}}.
\label{PEcurr}
\end{align}
These expression are used to obtain the plots in Figs.~\ref{diffcond}
and~\ref{noise}.
\\
\\
For charge degeneracy, $\Delta E_{C}=0$, the formula for the current
simplifies to $I^{th}=e(\Gamma_{01}^{+}-\Gamma_{01}^{-})/2$. This can
be used to derive Eqs.~\eqref{PEbelow}
and~\eqref{PEabove}. Considering temperatures such that only the first
mode has a finite population, the current below onset,
$I^{th}_{0}$, and above onset, $I^{th}_{1}$, are given in terms of
$\tilde{P}^{1}_{n}$ by
\begin{eqnarray}
  I^{th}_{0}&=&
  \frac{e^{2}V\tilde{\Gamma}}{4\hbar\omega_{1}}\left(\tilde{P}_{0}^{1}+2\sum_{n=1}^{\infty}\tilde{P}_{-n}^{1}\right)         ,\nonumber\\
  I^{th}_{1}&=&\frac{e^2V\tilde{\Gamma}}{4\hbar\omega_{1}}\left(\sum_{n=-1}^{1}\frac{eV-2n\hbar\omega_{1}}{eV}\tilde{P}^{1}_{n}+2\sum_{n=2}^{\infty}\tilde{P}_{-n}^{1}\right). \nonumber \\
\label{thcurr}
\end{eqnarray}
For temperatures $k_{B}T_{ph}\ll \hbar\omega_{1}$ we have
$\tilde{P}^{1}_{0}=1, \tilde{P}^{1}_{1}=\lambda_{1}^{2}$ and
$\tilde{P}^{1}_{n}$ with $n\leq -1$ exponentially suppressed. Then
Eq.~\eqref{thcurr} gives Eqs.~\eqref{PEbelow} and~\eqref{PEabove}. We
also note that from Eq. \eqref{thcurr} we have 
\begin{eqnarray}
 && \frac{\Delta
    G^{th}}{G^{th}_{0}}=\frac{\tilde{P}^{1}_{1}-\tilde{P}^{1}_{-1}}{\tilde{P}^{1}_{0}+
    2\sum_{n=1}^{\infty}\tilde{P}_{-n}^1}\leq \frac{\tilde{P}^{1}_{1}-\tilde{P}^{1}_{-1}}{\tilde{P}^{1}_{0}}  \nonumber\\
  &=&\frac{2\sinh\left(\frac{\hbar\omega_{1}}{2k_{B}T_{ph}}\right)I_{1}\left[\lambda_{1}^{2}/\sinh(\frac{\hbar\omega_{1}}{2k_{B}T_{ph}})\right]}{I_{0}\left[\lambda_{1}^{2}/\sinh(\frac{\hbar\omega_{1}}{2k_{B}T_{ph}})\right]}\leq
  \lambda_{1}^{2},
\label{PEcondstep}
\end{eqnarray}
where $\Delta G^{th}=G_1^{th}-G_0^{th}$. The conductance
step $\Delta G^{th}/G_0^{th}$ is thus limited above by
$\lambda_{1}^{2}$.  \\ \\ For charge degeneracy, $\Delta E_{C}=0$, the
expression for the noise simplifies to
$S^{th}=e^{2}(\Gamma_{01}^{+}+\Gamma_{01}^{-})/4$. For temperatures
such that only the first mode has a finite population the noise below, $S_0^{th}$,
and above, $S_1^{th}$, onset can be written as
\begin{eqnarray}
S^{th}_{0}&=&\frac{e^{2}}{8\hbar\omega_{1}}\left(eV\tilde P_{0}^{1}+\sum_{n=1}^{\infty}4n\hbar\omega_{1}\tilde{P}^{1}_{-n}\right),
\nonumber\\
S^{th}_{1}&=&\frac{e^{2}}{8\hbar\omega_{1}}\left(\sum_{n=-1}^{1}(eV-2n\hbar\omega_{1})\tilde P^{1}_{n}+\sum_{n=2}^{\infty}4n\hbar\omega_{1}\tilde{P}^{1}_{-n}\right). \nonumber \\
\label{thnoise}
\end{eqnarray}
For temperatures $k_{B}T_{ph}\ll \hbar\omega_{1}$ these formulas
reduce to Eqs.~\eqref{Bonsetnoise} and~\eqref{lowlimitnoise}. It is
clear from Eqs. (\ref{thnoise}) and (\ref{thcurr}) that the thermal
Fano-factors $F^{th}_{0}=S^{th}_{0}/(eI^{th}_{0})$ and
$F^{th}_{1}=S^{th}_{1}/(eI^{th}_{1})$ decreases monotonically with
bias voltage and that $F^{th}_{1}<F^{th}_{0}$. Hence, the Fano-factor
decreases monotonically with bias voltage.


\begin{thebibliography}{87}
\bibitem[{not()}]{note1}
\bibinfo{note}{For models in molecular electronics higher order, anharmonic
  terms in the phonon modes effectively make the charging energy positive for
  sufficiently many electrons on the dot, for arbitrary strong electron-phonon
  coupling. Such anharmonic terms are not present in our system, where the
  exact Hamiltonian is quadratic in all variables.}
\end{thebibliography}
\end{document}